\documentclass[12pt]{article}

\usepackage{jheppub}
\makeatletter
\def\@fpheader{\relax}
\makeatother

\usepackage{amsthm,amsmath,amssymb,mathtools}
\usepackage{mathrsfs}
\usepackage{bm}
\usepackage{bbm}
\usepackage{graphicx}
\usepackage{float}
\usepackage{tabularx}
\usepackage{caption}
\usepackage{subcaption}
\usepackage[most]{tcolorbox}
\usepackage[mathscr]{euscript}
\usepackage{dsfont}
\usepackage[mathscr]{euscript}
\usepackage{mathrsfs}
\usepackage{hyperref}
\usepackage{tikz}

\newcommand\blfootnote[1]{%
	\begingroup 
	\renewcommand\thefootnote{}\footnote{#1}%
	\addtocounter{footnote}{-1}%
	\endgroup 
}
\newenvironment{nohyphens}{%
	\hyphenpenalty=10000
	\exhyphenpenalty=10000
	\sloppy %
}{\par}

\title{Probing decoupled Throats of AdS$_{D}$ Black Holes in $D=6,7$}

\author[a]{Weichao Bu}
\author[a]{\!\!, Yang Lei\blfootnote{*The authors are ordered purely alphabetically and should all be viewed as the co-first authors. }}

\affiliation[\,a]{Institute for Advanced Study \& School of Physical Science and Technology,  Soochow University, No.1 Shizi Street, Suzhou 215006, P.R.~China}

\emailAdd{ring6030asdfghj@gmail.com}
\emailAdd{leiyang@suda.edu.cn}

\abstract{\begin{nohyphens}
The Kerr/CFT correspondence establishes a relationship between extremal black holes in higher dimensions and a chiral conformal field theory (CFT) in their near-horizon limit. 
A generalization of this framework, known as the EVH/CFT correspondence, has been developed for four- and five-dimensional AdS black holes. 
It was further proposed in \cite{Goldstein:2019gpz} that a generalized duality between $(D-2)$-dimensional geometry and $(D-3)$-dimensional field theory may emerge in AdS$_{D=6,7}$ black holes under a suitably defined extremal vanishing horizon (EVH) limit. 
In this work, we demonstrate that the near-EVH geometries arising in these AdS$_{6,7}$ black hole models, under the EVH limit, reduce to lower-dimensional black hole solutions whose metrics are conformally related to configurations of Einstein-Maxwell-Maxwell-dilaton (EMMD) gravity. 
This structural resemblance suggests a potential route toward a microscopic counting of non-AdS black hole entropy via higher-dimensional AdS/CFT techniques.

	\end{nohyphens}
}

\date{}

\begin{document} 
	
	\maketitle
	
\section{Introduction}\label{sec:intro}
So far, all gravitational-wave observations of black hole mergers have been consistent with the Kerr model of black holes \cite{LIGOScientific:2025wao}.
Understanding the quantum effects associated with Kerr black holes is therefore crucial for uncovering potential signatures of quantum gravity.
One of the major observational milestones to date is the confirmation of Hawking’s area theorem \cite{Hawking:1971vc}, which states that the total event-horizon area of black holes cannot decrease over time.
In particular, recent observations verify that the final horizon area after a merger exceeds the sum of the areas of the two initial black holes \cite{Tang:2025jyj,KAGRA:2025oiz}.

The problem of understanding the microscopic states of black holes is a cornerstone test for any candidate theory of quantum gravity.
Two main approaches have been developed to study this problem in different frameworks.
\begin{itemize}
\item The AdS$_3$/CFT$_2$ correspondence provides a particularly transparent setting, as the powerful Virasoro symmetry completely determines the microscopic states of black holes in the near-horizon region \cite{Strominger:1997eq}.
This idea was later generalized to higher-dimensional extremal black holes through the Kerr/CFT correspondence \cite{Guica:2008mu} (also generalized as the extremal black hole/CFT correspondence \cite{Lu:2009gj,Chow:2008dp,Lu:2008jk}).
These theories assert that the generic AdS$_2$ geometry emerging in the near-horizon limit of extremal black holes admits boundary conditions under which a Virasoro algebra governs the dynamics, without acknowledging the details of the higher dimensional black hole in the UV\footnote{There are also extensions of this idea to non-extremal black holes, where the Virasoro symmetry is realized as a geometric symmetry of the phase space of perturbations \cite{Frolov:2017kze}.}.
\item Within the  AdS$_{d+1}$/CFT$_d$ ($d=D-1>2$) framework, the microscopic states of AdS black holes can be reproduced by counting gauge-invariant operators in the dual superconformal field theories.
Remarkable progress has been achieved in understanding black holes in AdS$_4$ \cite{Kinney:2005ej, Benini:2015eyy,Choi:2018fdc,Choi:2019zpz,Choi:2019dfu}, AdS$_5$ \cite{Cabo-Bizet:2018ehj,Benini:2018ywd,Choi:2018hmj,Goldstein:2020yvj}, AdS$_6$ \cite{Choi:2019miv,Choi:2018fdc}, and AdS$_7$ \cite{Choi:2018hmj,Nahmgoong:2019hko,Bobev:2025xan,Hosseini:2018dob}.
See \cite{Zaffaroni:2019dhb} for a recent review. 
Most of these studies focus on BPS black holes, for which the counting of gauge-invariant states in the dual weakly coupled field theory remains valid.
An exception is the work \cite{Larsen:2019oll}, which investigates the field computation to understand the entropy of the near-BPS AdS$_5$ black holes.
\end{itemize}

The two approaches were investigated in a unified framework within the BPS black holes in the AdS$_D$ spacetimes. 
There are two possible perspetives being considered in the literature.
One is to consider the gravitational Cardy limits \cite{David:2020ems} where the charges and angular momenta are rescaled to make the black hole ultra spinning. 
The geometry in these limits are generically AdS$_2$ and AdS$_3$ \cite{Sadeghian:2014tsa,Kunduri:2007vf}, where Virasoro algebra can determine the dynamics in the near horizon limit. 
Another is building on earlier studies of the Extremal Vanishing Horizon (EVH) limit of black holes \cite{Balasubramanian:2007bs,deBoer:2011zt,Johnstone:2013eg,Fareghbal:2008eh,Fareghbal:2008ar,Sheikh-Jabbari:2011sar,Yavartanoo:2012yua, Yavartanoo:2012zz, Yavartanoo:2012zza,Noorbakhsh:2017nde,Sadeghian:2015laa,Sadeghian:2015hja} and explore the AdS$_5$ black holes in the BPS-EVH limits \cite{Goldstein:2019gpz}.
In the EVH limit, the black hole area $S_0\sim A$ is taken to zero while the central charge $c$ diverges, such that the product $c A$ remains fixed.
Given the expansion of entropy $S$ in terms of temperature $T$ is 
\begin{equation}\label{eq:S-Trelation}
	S(T,Q_i,J_a) = S_0(Q_i,J_a)+ S_1(Q_i,J_a) T+  S_2(Q_i,J_a)T^2+ \cdots \,,
\end{equation} 
the dynamics of the extremal black hole can be activated in this limit by taking $S_0 =0$, thereby circumventing the instability problem associated with AdS$_2$ geometry \cite{Maldacena:1998uz}, and the geometry is elevated to AdS$_3$ if $S_1 \neq 0$.
In the supersymmetric AdS$_5$ black hole model, an extremal pinched BTZ black hole emerges in the near-EVH regime, consistent with the $S\sim T$ relation \cite{Goldstein:2019gpz,deBoer:2011zt,Johnstone:2013eg, Sheikh-Jabbari:2011sar,Balasubramanian:2007bs,Fareghbal:2008ar,Yavartanoo:2012zz,Yavartanoo:2012yua}.
Moreover, since the microscopic states of the AdS$_5$ black hole can be computed via the superconformal indices of $\mathcal{N}=4$ SYM, the authors of \cite{Goldstein:2019gpz} demonstrated that a Cardy-like formula—used to count the degeneracy of an emergent CFT$_2$ \cite{Cardy:1986ie}—can be derived from the inverse Laplace transformation of the $\mathcal{N}=4$ superconformal indices.
This work represents a significant step toward proving the Kerr/CFT correspondence from the perspective of AdS/CFT, at least in the BPS limit. 
However, neither of these works can be considered as completing the proof of Kerr/CFT from AdS/CFT, as the mechanism of the emergent Virasoro algebra is still unclear. 

An interesting generalization of EVH black holes proposed in \cite{Goldstein:2019gpz,Balasubramanian:2007bs,deBoer:2011zt,Johnstone:2013eg,Fareghbal:2008eh,Fareghbal:2008ar,Sheikh-Jabbari:2011sar,Yavartanoo:2012yua, Yavartanoo:2012zz, Yavartanoo:2012zza,Noorbakhsh:2017nde,Sadeghian:2015laa,Sadeghian:2015hja} is the emergent EVH configurations in the IR limit of AdS$_6$ \cite{Chow:2008ip} and AdS$_7$ black holes \cite{Chow:2007ts,Bobev:2023bxl}. 
These black holes can be appropriately embedded in the ten dimensional string theory and eleven dimensional M-theory \cite{Cvetic:1999xp,Duff:1999rk}.
The extremal limit of higher dimensional AdS black holes requires more charges and angular momenta to support \cite{Gibbons:2004js,Chen:2006xh,Lu:2009gj}. 
This is due to the fact that the gravitational interactions of higher dimensions are stronger which requires more gauge forces or rotations to balance to get extremal black holes. 

On the other hand, when more charges and spatial dimensions are present, we might expect \eqref{eq:S-Trelation} can have higher orders of scalings $S\sim T^2$ or $S\sim T^3$ by fine tuning the charges and angular momenta i.e. make $S_0(Q_i,J_a)= S_1(Q_i,J_a)=0$ \cite{deBoer:2011zt,Johnstone:2013eg,Sheikh-Jabbari:2011sar}. 
However, we will clarify that near-EVH black holes in rotating AdS$_6$ and AdS$_7$ can be defined without being restricted to the near-BPS limit, as previously proposed in \cite{Goldstein:2019gpz}, and can instead be realized through general near-extremal limits.
We carefully examine the scaling relations between entropy $S$ and temperature $T$ for these black holes. 
Our analysis shows that their near-horizon geometries, both in $D=6$ and $7$, reduce to lower-dimensional black holes whose metrics exhibit a conformal equivalence to those of Einstein–Maxwell–Maxwell–Dilaton (EMMD) gravity, suggesting a close connection.
These black holes emergent from EVH limits are no longer with AdS asymptotics. 
However, as the microscopic states of the supersymmetric AdS$_{6,7}$ black holes can be reproduced from the dual SCFT, we should in principle to use the higher dimensional SCFT to understand the quantum states dual to these EMMD black holes. 
This can be an essential step to understand holographic duals to non-AdS black holes. 

The paper is organized as follows. In Section \ref{sec:reviewADs5}, we review the concept of near-EVH limits for Kerr–AdS$_5$ black holes based on \cite{Johnstone:2013eg,Goldstein:2019gpz,Sheikh-Jabbari:2011sar}. In Sections \ref{sec:ADs6} and \ref{sec:ADs7}, we investigate the near-EVH phases in AdS$_6$ and AdS$_7$ black holes, respectively, within this geometric framework. Specifically, for AdS$_6$, the near-EVH geometry corresponds to a four-dimensional EMMD black hole with $S \sim T^2$. In AdS$_7$, we find both $S \sim T$ and $S \sim T^3$ configurations, which are respectively realized as a BTZ black hole and a five-dimensional EMMD black hole. Finally, in Section \ref{sec:holography}, we discuss the holographic implications of these near-EVH limits, and conclude with discussions in Section \ref{sec:discussion}.

\section{Review of AdS$_5$}\label{sec:reviewADs5}

Studies on the (near)-EVH limits of AdS$_5$ black holes were initiated in \cite{Balasubramanian:2007bs,Fareghbal:2008ar} for static R-charged black holes and further developed in \cite{deBoer:2011zt}.
Subsequent investigations into the EVH limits of Kerr–RN–AdS$_5$ black holes and their dual field theory interpretations began with \cite{Johnstone:2013eg}, while an index-based interpretation of the emergent IR CFT$_2$ and its entropy was later provided in \cite{Goldstein:2019gpz}.
Generic AdS$_5$ black holes in U$(1)^3$ gauged supergravity carry three charges and two angular momenta.
The most general non-supersymmetric solution of this type was constructed in \cite{Wu:2011gq}, building on earlier works \cite{Gutowski:2004yv,Cvetic:2004ny,Chong:2005da,Chong:2005hr,Kunduri:2006ek}.
However, the full generality of these solutions makes them rather complicated. For the purpose of capturing the essential physics of near-EVH limits and their holographic interpretation via superconformal indices \cite{Goldstein:2019gpz}, the simpler special solutions presented in \cite{Chong:2005da,Johnstone:2013eg} suffice.
We therefore restrict our analysis to these special cases.

The black hole solution under consideration, studied in \cite{Lu:2008jk,Goldstein:2019gpz}, carries two equal charges and is described by the following five-dimensional metric \cite{Chong:2005da}:
\begin{equation}\label{eq:nonextremalBH}
	\begin{aligned}
		ds^2_5 &= H^{-\frac{4}{3}} \left[ -\frac{X}{\rho^2} (dt-a\sin^2 \theta \frac{d\phi}{\Xi_a}-b \cos^2 \theta \frac{d\psi}{\Xi_b})^2  + \frac{C}{\rho^2} \left( \frac{a b}{f_3} dt- \frac{b}{f_2} \sin^2 \theta \frac{d\phi}{\Xi_a} -\frac{a}{f_1} \cos^2 \theta \frac{d\psi}{\Xi_b}\right)^2 \right. \\ 
		&+ \left. \frac{Z \sin^2\theta}{ \rho^2} \left(\frac{a}{f_3} dt-\frac{1}{f_2} \frac{d\phi}{\Xi_a}\right)^2+ \frac{W \cos^2\theta}{\rho^2} \left(\frac{b}{f_3}dt-\frac{1}{f_1} \frac{d\psi}{\Xi_b}\right)^2 \right] + H^{\frac{2}{3}} \left[ \frac{\rho^2}{X} dr^2 + \frac{\rho^2}{\Delta_\theta} d\theta^2 \right],\\ 
		H&= 1+ \frac{q}{\rho^2},\quad \rho^2 =r^2 +a^2\cos^2 \theta +b^2 \sin^2 \theta , \quad 	\Delta_\theta = 1-a^2 \cos^2\theta -b^2  \sin^2 \theta\\
		f_1&= a^2+r^2, \quad f_2 =b^2+r^2, \quad f_3 =(a^2+r^2 )(b^2+r^2) +qr^2 ,\\ 
 X&= \frac{(a^2+r^2)(b^2+r^2)}{r^2} -2m +(a^2+r^2+q) (b^2+r^2+q) ,\\
		C &= f_1f_2(X+2m-\frac{q^2}{\rho^2}),  \qquad \Xi_a= 1-a^2, \quad \Xi_b =1-b^2,  \\ 
		Z&= -b^2 C +\frac{f_2f_3}{r^2}\left[f_3 - r^2(a^2-b^2)(a^2+r^2+q)\cos^2 \theta \right] ,\\ 
		W&= -a^2 C +\frac{f_1f_3}{r^2}\left[f_3 + r^2(a^2-b^2)(b^2+r^2+q)\sin^2 \theta \right] .
	\end{aligned}
\end{equation}
The corresponding thermodynamic quantities are given as follows:
\begin{eqnarray}\nonumber
\Omega_a&=& \frac{a(r_+^4+r_+^2b^2+r_+^2 q+b^2+r_+^2)}{(r_+^2 +a^2)(r_+^2+b^2)+q r_+^2}, \qquad \Omega_b=  \frac{b(r_+^4+r_+^2a^2+r_+^2 q+a^2+r_+^2)}{(r_+^2 +a^2)(r_+^2+b^2)+q r_+^2} , \\  \nonumber
\Phi_1&=& \Phi_2 = \frac{\sqrt{q^2+2m q}r_+^2}{(r_+^2+a^2)(r_+^2+b^2)+qr_+^2}, \qquad \Phi_3 = \frac{q ab}{(r_+^2+a^2)(r_+^2+b^2)+qr_+^2}\,, \\ \label{eq:generalchemicalpotential}
J_a &=& \frac{\pi a(2m +q\Xi_b)}{4G_N \Xi_b\Xi_a^2} , \qquad J_b =  \frac{\pi b(2m +q\Xi_a)}{4G_N \Xi_a\Xi_b^2}, \\  \nonumber
Q_1 &=& Q_2 =\frac{\pi \sqrt{q^2+2mq}}{4G_N \Xi_a\Xi_b} , \qquad Q_3 = -\frac{\pi a b q}{4G_N \Xi_a\Xi_b}, \\  \nonumber
S&=& \frac{ \pi^2 [(r_+^2+a^2)(r_+^2+b^2)+q r_+^2]}{2G_N \Xi_a \Xi_b r_+},\quad T =  \frac{2r_+^6 +r_+^4 (1+a^2+b^2+2q)-a^2b^2}{2\pi r_+ [(r_+^2+a^2)(r_+^2+b^2)+q r_+^2]}, \\  \nonumber
E&=& \frac{\pi}{8G_N \Xi_a^2 \Xi_b^2}[2m(2\Xi_a+2\Xi_b-\Xi_a\Xi_b)+q (2\Xi_a^2+2\Xi_b^2+2\Xi_a \Xi_b-\Xi_a^2\Xi_b-\Xi_b^2 \Xi_a)]  \,.
\end{eqnarray}
This solution is parametrized by four independent parameters $(a,b,m,q)$ and we have set the cosmological constant $g=1$. 
Here Newton constant $G$ is related to the rank of SU$(N)$ gauge group of $\mathcal{N}=4$ SYM by $\frac{\pi}{2G_N} = N^2$.
These expressions fully characterize the thermodynamic state of the black hole. The solution and its thermodynamic data play a central role in the subsequent analysis of the near-EVH limit and the emergence of an effective two-dimensional conformal description.

The EVH and near-EVH limits are defined as \cite{Johnstone:2013eg} \footnote{The EVH limit is defined as $b=r_+=0$ in \cite{Goldstein:2019gpz} which is still valid in $q=0$ limit.
Due to the symmetry between $a,b$ parameters, this will not change the essential physics.} 
\begin{equation}\label{eq:EVHcondition-AdS5}
\text{EVH}:	a=r_+=0 \,; \qquad \text{near-EVH}:\,  a=\lambda \epsilon^2, \quad r  = \epsilon x  \,,
\end{equation}
where $\epsilon \to 0$ is taken as also the near horizon limits. 
The EVH is treated as the ground state while the near-EVH limit can be treated as excited states of the theory, as it has non-vanishing temperature of order $\epsilon$. 
The corresponding geometries of (near)-EVH limits are respectively pinched AdS$_3$ and BTZ black holes. 
The entropy of these BTZ black hole scales as $S\sim N^2 \epsilon$, which is finite as we keep $N^2 \epsilon$ fixed as $N \to\infty$.
For the classical description of gravity to be valid, $N^2\epsilon$ should be taken large. 

We are especially interested in the near-EVH limit combined with the BPS condition, which includes both the supersymmetry 
\begin{equation}\label{eq:AdS5-supersymmetry}
	E= J_a+ J_b +Q_1 +Q_2 +Q_3 \,,
\end{equation}
and the extremality condition which requires horizons being degenerate.
The chemical potentials satisfying the supersymmetry condition are generically complex, and it is useful to define following chemical potentials:
\begin{equation}\label{eq:def-chemicalpotential}
	\Delta_i = \beta(1-\Phi_i), \qquad \omega_I= \beta(1-\Omega_I)\,, \qquad i=1,2,3; \quad  I=1,2  \,,
\end{equation}
and are subject to the linear constraint:
\begin{equation}
	\Delta_1+\Delta_2+\Delta_3-\omega_a-\omega_b =2\pi i\,. 
\end{equation}
This makes the parameters $(q,m)$ generically complex unless the radius of horizon is determined by
\begin{equation}\label{eq:abr0-AdS5}
	r_0^2 = \frac{ab}{1+a+b}\,.
\end{equation}
This condition removes the closed timelike curves in the spacetime.
The solutions satisfying both the supersymmetry condition \eqref{eq:AdS5-supersymmetry} and the horizon size condition \eqref{eq:abr0-AdS5} have real values of charges and entropy, parametrized by the $(q,m)$ parameters as
\begin{align}\label{eq:BPSmqr0}
	\begin{split}
 q=  \frac{(a+b)(1+a)(1+b)}{1+a+b}\,, \quad  m =  \frac{(a+b)^2(1+a)(1+b)(2+a+b)}{2(1+a+b)} \,.
	\end{split}
\end{align}
Combining the BPS conditions \eqref{eq:AdS5-supersymmetry} and  \eqref{eq:abr0-AdS5} and the EVH conditions \eqref{eq:EVHcondition-AdS5}, the entropy of the black hole is determined to be
\begin{equation}\label{eq:AdS5-EVH-BPS-S}
	S = \frac{\pi  b}{1-b} \sqrt{\frac{b \lambda}{1+b}} N^2\epsilon \,.
\end{equation}
The corresponding decoupling metric of AdS$_5$ is of the following form 
\begin{align}\label{eq:EVHnearhorizon}
	\begin{split}
		ds^2 &=\left(\frac{h}{\sin \theta}\right)^{-\frac{4}{3}} \left[ h^2 ds_3^2 	  + h^2\frac{b^2 d\theta^2}{1-b^2\sin^2\theta} + \frac{1-b^2\sin^2\theta}{(1-b)^2} \frac{\cos^2\theta}{\sin^2\theta} d\tilde{\psi}^2\right] \,, \\
& h = \sin \theta + \frac{1}{b\sin \theta}\,, \qquad \tilde{\psi} = \psi-(1-b )t \,,
	\end{split}
\end{align}
with $ds_3^2$ in the EVH limit is taken as the metric of AdS$_3$
\begin{equation}
	ds_3^2 = - \frac{x^2}{\ell_3^2} d\tau^2 + \frac{\ell^2_3}{x^2}dx^2+x^2d\widetilde{\chi}^2, \qquad \ell_3 = \frac{b}{1+b}, \quad t= \frac{1+b }{b} \tau \,,
\end{equation}
while in the near-EVH case, the $ds_3^2$ is replaced by the extremal BTZ geometry: 
\begin{equation}\label{eq:near-EVHBTZ}
	ds_{3}^2 = - \frac{\left(x^2-x_0^2\right)^2}{\ell_3^2 x^2} d\tau^2 + \frac{\ell_3^2x^2 dx^2}{(x^2-x_0^2)^2}  +x^2  \left(d\widetilde{\chi}-\frac{x_0^2}{\ell_3 x^2}d\tau\right)^2 \,.
\end{equation}
This geometry is called pinched as the periodicity of $S^1$ direction $\tilde{\chi}= \epsilon \phi$ is $2\pi \epsilon$. 

We are now ready to explain the entropy \eqref{eq:AdS5-EVH-BPS-S} of the BPS-EVH black hole from the computations of the dual $\mathcal{N}=4$ SYM  \eqref{eq:near-EVHBTZ}, following \cite{Goldstein:2019gpz}.
It has been shown in various works
\cite{Hosseini:2017mds,Choi:2018hmj,Cabo-Bizet:2018ehj,Benini:2018ywd,Goldstein:2020yvj} that the entropy functional in the large $N$ limit of $\mathcal{N}=4$  SYM is
\begin{eqnarray}
\nonumber
S &=& \ln Z + (J_a+Q_3) \omega_a + (J_b+Q_3) \omega_b + (Q_1-Q_3) \Delta_1 +(Q_2-Q_3) \Delta_2  +2\pi i Q_3  \\
\ln Z & =& \frac{N^2}{2} \frac{\Delta_1\Delta_2\Delta_3}{\omega_a \omega_b}, \qquad \Delta_1 +\Delta_2+\Delta_3 -\omega_1 -\omega_b =2\pi i \,,
\label{eq:AdS5-S-lnZ}
\end{eqnarray}
where the chemical potentials are taken to be complex to avoid cancellations between bosonic and fermionic degrees of freedom \cite{Kinney:2005ej}. 
After performing a saddle point approximation on the chemical potentials $\Delta_i,\omega_a,\omega_b$, the entropy functional \eqref{eq:AdS5-S-lnZ} reproduces the BPS black hole entropy \cite{Kunduri:2006ek}
\begin{equation}\label{eq:AdS5-S}
	S= 2\pi \sqrt{Q_1 Q_2 +Q_1 Q_3 +Q_2 Q_3 - \frac{N^2}{2} (J_a +J_b)} \,.
\end{equation}
The entropy formula \eqref{eq:AdS5-S} is consistent with the one for the two equal-charge black hole \eqref{eq:generalchemicalpotential} in the BPS limit. 
Therefore, the BPS-EVH limit of the entropy \eqref{eq:AdS5-EVH-BPS-S} should be also encoded in the superconformal indices of the $\mathcal{N}=4$ SYM, and the entropy functional \eqref{eq:AdS5-S-lnZ}.

In fact, as shown in \cite{Goldstein:2019gpz}, the near-EVH limit splits the extremization of the entropy functional in the saddle-point approximation into two steps.
In the large-$N$ limit with the EVH condition $N^2\epsilon$ held fixed, the charges scale as $Q_3\sim J_a\sim N^2\epsilon^2$, while $Q_{1,2}\sim J_b\sim N^2$.
Under this scaling, the saddle-point approximation of the functional in \eqref{eq:AdS5-S-lnZ} is justified only for $\Delta_{1,2}$ and $\omega_b$, but not for $\omega_a$.
Therefore, setting $\Delta_1 = \Delta_2 = \Delta$ in the solution \eqref{eq:nonextremalBH} and performing the $\omega_b$ and $\Delta$ integrations first, we obtain
\begin{equation}\label{eq:secondstep-saddle}
	e^S = \int d\omega_a \exp\left[
	\frac{N^2 \hat{\Delta}^2}{ 2 \omega_a} \left( \frac{\hat{\Delta}}{ \hat{\omega}_b} - 1 \right) + 2\pi i Q_3
	\right] e^{\omega_a(J_a + Q_3)},
\end{equation}
where $\hat{\Delta}$ and $\hat{\omega}_b$ denote the values of the chemical potentials satisfying the saddle-point equations.
This expression \eqref{eq:secondstep-saddle} closely resembles the Cardy formula of a two-dimensional CFT. 
It may be viewed as a functional over the rescaled modulus $\tilde{\omega}_a = \epsilon \omega_a$.
Since $N^2 \epsilon$ is fixed and large, extremization over $\tilde{\omega}_a$ is valid and reproduces precisely the near-EVH entropy given in \eqref{eq:AdS5-EVH-BPS-S}.

This computation strongly supports the emergence of an effective two-dimensional conformal field theory in the near-horizon limit.
The resulting EVH 2D CFT is closely related to the Kerr/CFT correspondence, though their central charges differ by a factor of $\sqrt{2}$ — a discrepancy that may stem from differing choices of time coordinates between AdS$_5$ and AdS$_3$ \cite{Goldstein:2019gpz,Johnstone:2013eg,Balasubramanian:2007bs,Fareghbal:2008ar}.
This result should therefore be viewed as a key first step toward uncovering the microscopic mechanism underlying the Kerr/CFT correspondence within the framework of AdS/CFT.

However, this does not constitute sufficient evidence to claim that the EVH/CFT correspondence (as a more special version of Kerr/CFT correspondence) has been derived from AdS/CFT.
First, although the central charge in the Cardy formula matches that of Kerr/CFT, its origin on the field theory side remains obscure. 
It is unclear how this central charge, which is geometrically defined by the AdS$_2$ throat in the gravity picture, can be derived from the algebraic data of $\mathcal{N}=4$ SYM, such as the conformal dimensions of operators or the central charge of $\mathcal{N}=4$ SYM.
Second, it is not understood how operators transforming under the 4D superconformal algebra PSU$(2,2|4)$ organize into representations of the Virasoro algebra\footnote{Evidence suggests that the emergence of the Virasoro algebra in the near-EVH limit is tied to the chiral algebra mechanism \cite{Goldstein:2019gpz}, since one of the constituents of the Virasoro generators $J_a + Q_3$ is a Schur operator. 
However, the set of letters contributing to the entropy and transforming under the Virasoro algebra also includes non-Schur operators. Therefore, a generalization of the chiral algebra mechanism appears necessary to account for the Virasoro symmetry emerging in near-EVH limits.}.
This conceptual tension is reflected in the mismatch between the local symmetries of AdS$_5$ in the UV and AdS$_3$ in the IR.
Despite these aspects, we are not aware of whether the general SL$(3,\mathbb{Z})$ families of AdS black hole corresponding to the root of unity configurations \cite{Cabo-Bizet:2019eaf,Benini:2020gjh,ArabiArdehali:2021nsx,Jejjala:2021hlt,Aharony:2021zkr,Cabo-Bizet:2021plf,Colombo:2021kbb,Aharony:2024ntg,Jejjala:2022lrm,Cabo-Bizet:2020ewf} can have such a near-EVH decoupling limit or not. 
These topics will not be discussed in this paper.

\section{AdS black holes in $D=6,7$}

AdS$_{D}$ black holes in $D=6$ and $7$ dimensions exhibit considerably richer structure than their four- and five-dimensional counterparts.
Of particular interest are those solutions that either possess known holographic duals or admit consistent embeddings into string theory.
Notable examples include AdS$_6$ black holes with two independent angular momenta and one R-charge \cite{Chow:2008ip}, as well as AdS$_7$ black holes carrying up to three angular momenta and two distinct R-charges \cite{Bobev:2023bxl}.
This classification closely mirrors the properties of superconformal field theories in $d=5$ and $6$ dimensions \cite{Nahm:1977tg}.
Moreover, such black hole solutions are of intrinsic interest due to their realizations as specific brane configurations in string theory or M-theory \cite{Cvetic:1999xp,Duff:1999rk}.

The metric of AdS$_{D}$ is expressed in coordinates including time $t$, a radial coordinate $r$, along with $[\frac{D-1}{2}]$ azimuthal angles $\phi_i$ and $[\frac{D-2}{2}]$ latitude coordinates $y_\alpha$ on the sphere.
To be precise, we set $D = 2n + 1$ for odd dimensions and $D = 2n$ for even dimensions. 
The coordinates $y_\alpha$ are related to the direction cosines $\mu_i$ of the unit sphere $S^{[D/2]}$ via the Jacobi transformation:
\begin{equation}\label{eq:Jacobi-transformation}
	\mu_i^2 = \frac{ \prod_{\alpha=1}^{n-1} (a_i^2 - y_\alpha^2)}{\prod_{k \neq i}^{n}(a_i^2 - a_k^2)}, \qquad i = 1, \dots, \left[\frac{D-1}{2}\right],
\end{equation}
which automatically satisfies the constraint $\sum_{i=1}^{[D/2]} \mu_i^2 = 1$. 
Here, each $a_i$ parametrizes a rotation in the corresponding $\phi_i$ direction.
The symbol $\prod'$ indicates that the product omits any vanishing factor. In the case of even $D$, we set $a_n = 0$.
This coordinate system, introduced in \cite{Chen:2006xh}, provides a natural framework for generalizing Kerr–AdS black holes to include NUT charges. It also offers several structural advantages: for example, the metric on $S^{D-2}$ becomes diagonal in these coordinates, and the coordinates $y_\alpha$ and $r$ appear in a highly symmetric manner throughout the metric.

In this section, we will describe the corresponding (near)-EVH limits of the AdS$_{D}$ black holes with $D=6,7$. 
The analysis on the thermodynamics of these black holes indicate they have $S\sim T^{D-4}$ scaling relations in the near-BPS and near-EVH limits \cite{Goldstein:2019gpz}.
These scaling relations between entropy and temperature was conjectured to hint a possible AdS$_{D-2}$/CFT$_{D-3}$ duality emergent in the IR. 
In this section, we will mainly make two clarifications in these models. 
\begin{itemize}
	\item The near-BPS condition is not necessary to define the near-EVH limits.
\item The near-EVH geometry is not of the AdS$_{D-2}$ black holes, but instead the black holes displaying structures of solutions to EMMD theories \cite{Lu:2013eoa} in $(D-2)$-dimensional manifold.
\end{itemize}

\subsection{AdS$_{6}$ black hole}\label{sec:ADs6}
The AdS$_6$ black hole is solution to the $\mathcal{N}=4$, SU$(2)$ gauged supergravity theory in six dimensions, which includes a graviton, a two-form potential, a scalar, a one-form potential and together with the gauge potential of SU$(2)$ Yang-Mills theory. 
The bosonic part of Lagrangian is \cite{Cvetic:1999un,Chow:2008ip} 
\begin{eqnarray}\nonumber
\mathcal{L}_6 &=& R \star 1 - \frac{1}{2} \star d\varphi \wedge d\varphi - \frac{1}{2X^2} \Big(\star F_{(2)} \wedge F_{(2)} + \star F_{(2)}^I \wedge F_{(2)}^I \Big) \\ \label{eq:lagrangian-AdS6}
&& -\frac{1}{2}X^4 \star F_{(3)} \wedge F_{(3)} +\left( 9X^2 +\frac{12}{X^2} - \frac{1}{X^6}\right) \star 1 \\ \nonumber
&& -A_{(2)} \wedge  \left(
\frac{1}{2}dA_{(1)} \wedge dA_{(1)} + \frac{1}{\sqrt{2}} A_{(2)} \wedge dA_{(1)} + \frac{1}{3} A_{(2)} \wedge A_{(2)} + \frac{1}{2} F_{(2)}^I \wedge F_{(2)}^I
\right)\,.
\end{eqnarray}
In order to describe the AdS$_6$ black hole solutions of this theory conveniently, we make the following ansatz
\begin{equation}
	y_{\alpha} = (y,z), \quad (a_1,a_2,a_3)\equiv (a,b,0) \,.
\end{equation}
The charged rotating AdS$_6$ in the asymptotic static frame is \footnote{We take the static frame instead of other $\psi_i$ coordinates as these are appropriate for computing the thermodynamic quantities.
They are thus also suitable for taking the near-EVH limits later. 
} \cite{Chow:2008ip,Lu:2008jk}
\begin{align}\label{eq:AdS6-Kerr-RN}
	\begin{split}
		ds^2 &= H^{\frac{1}{2}} \left[ 
		\frac{(r^2+y^2)(r^2+z^2)}{X} dr^2 + \frac{(r^2+y^2)(y^2-z^2)}{Y} dy^2
		\right. \\
		&+\frac{(r^2+z^2)(z^2-y^2)}{Z} dz^2	- \frac{X}{H^2 (r^2+y^2)(r^2+z^2)} \mathcal{A}^2 \\
		& + \frac{Y}{(r^2+y^2)(y^2-z^2)} \left( (1+r^2) (1-z^2)  
		d\tilde{t}  - (a^2+r^2)(a^2-z^2) d\tilde{\phi}_1  \right. \\
		& \left.  -  (b^2+r^2)(b^2-z^2) d\tilde{\phi}_2 - \frac{q r \mathcal{A}}{ H(r^2+y^2)(r^2+z^2)}
		\right)^2  \\
		& + \frac{Z}{(r^2+z^2)(z^2-y^2)} \left( (1+r^2) (1-y^2)  
		d\tilde{t}  - (a^2+r^2)(a^2-y^2) d\tilde{\phi}_1  \right. \\
		&  \left. \left.  -  (b^2+r^2)(b^2-y^2) d\tilde{\phi}_2 - \frac{q r \mathcal{A}}{ H(r^2+y^2)(r^2+z^2)}
		\right)^2   \right]  \,,
	\end{split}
\end{align}
where the functions in the metric are explicitly
\begin{align}
	\begin{split}
		X&= (r^2+a^2)(r^2+b^2) + [r(r^2+a^2)+q]  [r(r^2+b^2)+q] -2m r \\ 
		Y& = -(1-y^2) (a^2-y^2)(b^2-y^2), \qquad Z= -(1-z^2) (a^2-z^2)(b^2-z^2) \\
		H & = 1+ \frac{q r}{(r^2+y^2)(r^2+z^2)}  \\
		\mathcal{A} &= (1-y^2) (1-z^2)  d\tilde{t} -(a^2-y^2)(a^2-z^2)  d\tilde{\phi}_1 -(b^2-y^2)(b^2-z^2)  d\tilde{\phi}_2 \,.
	\end{split}
\end{align}
Without loss of generality, we adopt the convention from \cite{Chow:2008ip} in which the coordinates $y,z$ are restricted to the region
\begin{equation}\label{eq:def-domainyz-AdS6}
	-a \le y\le a\le z\le b \,.
\end{equation}
The coordinates $\tilde{\phi}_i$ and $\tilde{t}$ are related to the standard Boyer-Linquist coordinates, where the $S^1$ directions $\phi_i$ have periodicity $2\pi$, via
\begin{equation}\label{eq:tilde-untilde-phi}
	\tilde{t} = \frac{t}{\Xi_a\Xi_b}, \qquad 
	\tilde{\phi}_1 = \frac{\phi_1}{a\Xi_a (a^2-b^2)} ,\qquad 	\tilde{\phi}_2 = \frac{\phi_2}{b\Xi_b (b^2-a^2)} \,.
\end{equation}
This redefinition is introduced to simplify the notation, following \cite{Chen:2006xh}. 
The thermodynamic quantities can be directly computed from the metric  \eqref{eq:AdS6-Kerr-RN}, yielding the following expressions \cite{Chow:2008ip,Choi:2018fdc,Mishra:2022gil,Wu:2021vch}
\begin{eqnarray} \nonumber
		E &=& \frac{2\pi m}{3G_N \Xi_a\Xi_b} \left[ \frac{1}{\Xi_a}+\frac{1}{\Xi_b} + \frac{q}{2m} \left(1+\frac{\Xi_b}{\Xi_a} + \frac{\Xi_a}{\Xi_b}\right)\right] ,  \quad 	S = \frac{2\pi^2[(r_+^2+a^2)(r_+^2+b^2)+q r_+]}{3G_N \Xi_a \Xi_b}\\ 
		\label{eq:thermo-AdS6}
		J_a &=& \frac{2\pi ma}{3G_N \Xi_a^2 \Xi_b} \left(1+\frac{
			\Xi_bq}{2m}\right), \qquad J_b = \frac{2\pi mb}{3G_N \Xi_a \Xi_b^2} \left(1+\frac{
			\Xi_aq}{2m}\right), \\ \nonumber
		T&=& \frac{2r_+^2(1+r_+^2)(2r_+^2+a^2+b^2)-(1-r_+^2)(r_+^2+a^2)(r_+^2+b^2) +4q r_+^3 -q^2}{4\pi r_+ [(r_+^2+a^2)(r_+^2+b^2)+q r_+]} \\ \nonumber
	\Omega_a &=& \frac{a[(r_+^2+1)(r_+^2+b^2) +q r_+]}{(r_+^2+a^2)(r_+^2+b^2)+q r_+}, \quad 
		\Omega_b = \frac{b[(r_+^2+1)(r_+^2+a^2) +q r_+]}{(r_+^2+a^2)(r_+^2+b^2)+q r_+}\\ \nonumber
	\Phi&=& \frac{ \sqrt{q^2+2m q} r_+ }{(r_+^2+a^2)(r_+^2+b^2)+q r_+}  \qquad Q= \frac{\pi \sqrt{q^2+2m q}}{G_N\Xi_a\Xi_b}\,.
\end{eqnarray}
The Newton constant $G_N$ is related to the field theory constant by $\frac{27\sqrt{2}}{5\pi} \frac{N^{\frac{5}{2}}}{\sqrt{8-N_f}} = G_N^{-1}$ \cite{Choi:2019miv}, where $N_f$ is the number of flavor symmetry. 
These quantities satisfy the first law of black hole thermodynamics \cite{Chow:2008ip}. 

We are particularly interested in the (near)-EVH limit, where the extremal black hole horizon scales as $\epsilon$ and $N^{\frac{5}{2}} \epsilon^2$  is held fixed to ensure the presence of non-trivial dynamics.
In this framework, the exact EVH limit corresponds to the ground state of the near-EVH geometry, characterized by a vanishing horizon.
Achieving such a configuration requires careful tuning of parameters, especially since in spacetime dimensions $D\ge 6$ , the existence of extremal AdS black holes generally demands additional charges and angular momenta to balance the gravitational potential.

The EVH limit is generically defined as 
\begin{equation}\label{eq:EVHlimits-AdS6}
a \equiv 0, \quad r= \epsilon x,  \quad r_+ =\epsilon x_+, \quad  t= \epsilon \tau, \quad q=0 \,.
\end{equation}
The blackening factor has a vanishing horizon by taking these parameters:
\begin{equation}
	X(r) = (r^2+b^2)(r^2+1)r^2 \,,
\end{equation}
indicating that the limit results in the ground state geometry. 
However, the function $Y$ and coordinate $y$ in the metric \eqref{eq:AdS6-Kerr-RN} vanish as the definition domain \eqref{eq:def-domainyz-AdS6}, while $\tilde{\phi}_1$ is singular in the EVH limit \eqref{eq:EVHlimits-AdS6}.
They thus need to be appropriately normalized. 
The solution adapted to both the definition domain \eqref{eq:def-domainyz-AdS6} and the Jacobi transformation \eqref{eq:Jacobi-transformation} in this EVH limit is: 
\begin{equation}\label{eq:EVH-singu-coordinate}
y=a\cos\theta_1, \quad z= b\cos \theta_2, \quad a\tilde{\phi_1} = -  \frac{\phi_1}{b^2}, \quad  0\le \theta_1 \le \pi,\qquad 0\le \theta_2 \le \frac{\pi}{2}\,.
\end{equation}
The geometry yields
\begin{eqnarray}\nonumber
ds^2 & = &- \frac{(1+r^2) (1-b^2 \cos^2\theta_2)}{1-b^2}  dt^2 +  \frac{r^2 + b^2 \cos^2\theta_2}{(r^2+1)(r^2+b^2)}  dr^2 + r^2 \cos^2\theta_2 (d\theta_1^2 + \sin^2\theta_1 d\phi_1^2) \\
		&& + \frac{r^2+b^2}{1-b^2} \sin^2\theta_2 d\phi_2^2
		+ \frac{r^2+b^2 \cos^2\theta_2}{1-b^2 \cos^2\theta_2} d\theta_2^2 \,. \label{eq:vacuum-AdS6-wthb}
\end{eqnarray}
A special limit is $z \to b$ where the geometry is manifestly decoupled as global AdS$_4$ spanned by $(t,r,\theta_1,\phi_1)$ with $S^2$ compact manifold spanned by $(\theta_2,\phi_2)$. 

We are now in a position to explore the near-EVH limit.
This limit is implemented via the following scalings
\begin{equation}\label{eq:near-EVH-AdS6-cond-1}
 a =\lambda \epsilon^2, \qquad q= bx_+  \epsilon  + q^{(3)} \epsilon^3 \,, \qquad  r_+= \epsilon x_+ + y_3 \epsilon^3 \,,
\end{equation}
together with the coordinate transformation \eqref{eq:EVH-singu-coordinate}
such that the range \eqref{eq:def-domainyz-AdS6} is preserved. 
The scaling between $a$ and $\epsilon$ was enlightened as possible generalizations of the near-EVH limits in the AdS$_5$ black hole counterparts \cite{Goldstein:2019gpz}. 
The $q^{(3)}$ has to be chosen to guarantee the existence of the extremal horizon, where the temperature vanishes. 
In order to see this, we introduce the $\epsilon^3$ order perturbation of radial coordinate in \eqref{eq:near-EVH-AdS6-cond-1}.
The temperature $T$ does not vanish at the order $\epsilon$ for generic values of $q^{(3)}$, which results in $S\sim T^2$ scaling relations: 
\begin{align}\label{eq:S-T-aDs6}
	\begin{split}
S &= \frac{2 \pi^2 b x_+^2}{3G_N (1-b) } \epsilon^2  \\
T& =  \frac{-2b x_+ q^{(3)} + (3+4b+3b^2)x_+^4 -b^2 \lambda^2 + 2y_3 b^2 x_+}{4\pi b(1+b)x_+^3} \epsilon + \mathcal{O}(\epsilon^3) \,.
	\end{split}
\end{align}
The temperature will vanish at the order $\epsilon$ if $q^{(3)}$ satisfies 
\begin{equation}\label{eq:q(3)-extremal}
	q^{(3)}=\left(2+\frac{3}{2b} + \frac{3b}{2}\right) x_+^3 - \frac{ b \lambda^2}{ 2x_+} +y_3b \,.
\end{equation}
More higher order of $\epsilon$ will not affect the expansion \eqref{eq:S-T-aDs6} and the extremality constraint \eqref{eq:q(3)-extremal}.
By repeating this expansion to fix the coefficient under the $\epsilon$ expansion, the black hole can be the decoupled near-EVH extremal black holes. 
For the decoupling conditions do not satisfy \eqref{eq:q(3)-extremal}, the near-EVH black holes are generically non-extremal.

Therefore, the near-EVH scaling relations are possible even without the near-BPS condition proposed in \cite{Goldstein:2019gpz}.
In the near-EVH limit, our interesting dynamics is restricted to 
\begin{equation}
	\frac{\epsilon^2}{G} \sim N^{\frac{5}{2}} \epsilon^2\quad  \text{fixed}, \qquad \epsilon\to 0 \,.
\end{equation}
Despite that the $S\sim T^2$ scaling relations is not that special EVH limit, the particular scaling of $a$ with 
$\epsilon$ was motivated by potential generalizations of near-EVH limits in the context of AdS$_5$ black holes \cite{Goldstein:2019gpz}.
Meanwhile, the scaling of 
$q$ with $\epsilon$ is chosen to ensure the persistence of an extremal horizon in the limit.
Moreover, this ansatz naturally incorporates near-BPS EVH limits without further adjustment.

The BPS limit of AdS$_6$ black hole and the field theory interpretation was discussed in \cite{Choi:2018fdc}.
The supersymmetry condition yields 
\begin{equation}
	E= J_a+J_b +Q \,.
\end{equation}
Similar to AdS$_5$ \cite{Goldstein:2019gpz,Cassani:2019mms}, the $q$ parameter satisfying the supersymmetry condition is generically complex unless the BPS values of horizon size $r_0$ is taken : 
\begin{equation}
	q= (a+b+ab)r_+-r_+^3 +i (1+a+b) \left(r_+^2 - r_0^2 \right), \qquad r_0^2 =\frac{a b}{1+a+b} \,.
\end{equation}
The chemical potentials defined at the BPS points are 
\begin{equation}
	\Delta=\lim_{r_+\to r_0} \beta(1-\Phi), \qquad \omega_a =\lim_{r_+\to r_0} \beta(1-\Omega_a), \qquad \omega_b =\lim_{r_+\to r_0} \beta(1-\Omega_b) \,,
\end{equation}
which are generically complex and subject to the condition
\begin{equation}
	\omega_a+\omega_b -3\Delta=2\pi i\,.
\end{equation}
This condition will be crucial to reproduce the black hole entropy from the superconformal indices, as will be discussed in section \ref{sec:holography}. 
As the generalization of EVH-BPS limit of AdS$_5$, the work \cite{Goldstein:2019gpz} considered the combined near-BPS and near-EVH limits of AdS$_6$ black hole, where the horizon size reduces to
\begin{equation}
x_+ = \sqrt{\frac{b \lambda}{1+b}} \,.
\end{equation}
Besides the entropy of the black hole in the BPS limit is
\begin{equation}\label{eq:near-EVH-bpsentropy}
	S= \frac{2\pi^2 b^2 \lambda}{3G_N (1-b^2)} \epsilon^2 \,,
\end{equation}
which is kept finite in the near-EVH limit. 
This is precisely a special case of near-extremal black hole \eqref{eq:S-T-aDs6}. 

The near-EVH metric with extremality corresponding to the condition \eqref{eq:near-EVH-AdS6-cond-1} is  \footnote{The temperature of extremal black hole must be zero. We need to choose $q^{(3)}$ in  \eqref{eq:near-EVH-AdS6-cond-1} appropriately to make $T$ identically vanishing. Although any choices will not affect the leading order in the near-EVH limit of the AdS$_6$ black holes.}: 
\begin{align}\label{eq:near-EVH-AdS6-2}
	\begin{split}
		d\tilde{s}^2 &= H^{\frac{1}{2}}\left[ \cos^2\theta_2 \epsilon^2  x^2 \frac{ dx^2}{ (x-x_+)^2} + \epsilon^2 \cos^2\theta_2 x^2d\theta_1^2
		+\epsilon^2 x^2 \cos^2\theta_2 \sin^2\theta_1 d\phi_1^2  \right. \\
		& - \frac{(x-x_+)^2 [b^2\sin^2\theta_2 d\phi_2 -\epsilon b(1-b^2 \cos^2\theta_2)d\tau]^2}{(x b \cos \theta_2+ x_+ \sec \theta_2 )^2 (1-b^2)^2} \\
		& \left. +\frac{b^2\cos^2\theta_2}{1-b^2\cos^2\theta_2} d\theta_2^2  +  \frac{b^2 \sin^2\theta_2 \cos^2\theta_2(1-b^2 \cos^2\theta_2) (b x + x_+)^2}{(1-b^2)^2( x_++xb \cos^2\theta_2)^2} d\phi_2^2  \right] \\
H& = 1+ \frac{ x_+}{x b \cos^2\theta_2} \,.
	\end{split}
\end{align}
The decoupled geometry behaves differently for different locations of $\theta_2$. 
We thus consider two limits respectively. 
In the $\theta_2 \to 0$ limit \eqref{eq:vacuum-AdS6-wthb}, 
the four dimensional black hole decoupled from the IR limit of  AdS$_6$ black hole is spanned by the coordinates $(\tau,x,\theta_1,\phi_1)$, whose metric takes the following ansatz
\begin{align}
	\begin{split}
		d\tilde{s}^2 =  H^{\frac{1}{2}}\left[ \epsilon^2 ds_4^2 +\frac{b^2}{1-b^2}d\Omega_2^2 \right]\,,   \qquad d\Omega_2^2=d\theta_2^2+ \sin^2\theta_2 d\phi_2^2  \,.
	\end{split}
\end{align}
The decoupled four dimensional black hole can be written as follows
\begin{equation}\label{eq:4d-decoupled-BH}
	ds_4^2 = - \frac{(x-x_+)^2}{(x+\frac{x_+}{b})^2} d\tau^2 + \frac{x^2 dx^2}{(x-x_+)^2} + x^2 d\theta_1^2+ x^2 \sin^2\theta_1 d\phi_1^2  \,.
\end{equation}
Typically, the metric \eqref{eq:4d-decoupled-BH} is conformal to the extremal EMMD black hole solutions \cite{Lu:2013eoa,Lu:2025eub}, which will also be reviewed in appendix \ref{appendix:EMMD}. 
For instance, the EMMD solution with $N_1=N_2=2$ is: 
\begin{equation}\label{eq:matching-EMMD-EVH}
ds^2_{\text{EMMD}} = \frac{\tilde{r}+q_2}{\tilde{r}+q_1} \left[ 
- \frac{(\tilde{r}-\mu)\tilde{r}}{ (\tilde{r}+q_2)^2} dt^2 + \frac{(\tilde{r}+q_1)^2}{\tilde{r}(\tilde{r}-\mu)} d\tilde{r}^2 + (\tilde{r}+q_1)^2 (d\theta_1^2 + \sin^2\theta_1 d\phi_1^2)
\right] \,.
\end{equation}
Without loss of generality, we assume $q_2>q_1$. 
The metric inside the square bracket of \eqref{eq:matching-EMMD-EVH} is precisely the decoupled near-EVH geometry \eqref{eq:4d-decoupled-BH} by identifying 
\begin{equation}\label{eq:AdS6-extremal-ID}
\mu=0, \quad x_+= q_1, \quad \frac{x_+}{b} = q_2-q_1, \quad  x= \tilde{r}+q_1\,.
\end{equation}
If the BPS condition is imposed \cite{Goldstein:2019gpz}, the horizon size $x_+$ and charge $q$ are both fixed by the rotation parameters, which will correspond to special values of two U$(1)$ charges in the EMMD theory. 

Recall that in AdS$_5$ black hole models, the EVH limit yields a pinched AdS$_3$ geometry in Poincar\'e coordinate, whereas the near-EVH limit leads to a pinched BTZ geometry \cite{Balasubramanian:2007bs,Goldstein:2019gpz,Johnstone:2013eg,deBoer:2011zt}. 
This can be understood holographically: the BTZ geometry is dual to thermal excitations above the AdS$_3$ vacuum. 
The situation in the AdS$_6$ black hole model studied here is qualitatively different. In the exact EVH limit, the geometry described by \eqref{eq:vacuum-AdS6-wthb} contains a decoupled global AdS$_4$ submanifold, whose $r\to 0$ limit reduces to a four-dimensional Minkowski spacetime. The decoupled black hole geometry \eqref{eq:4d-decoupled-BH} can therefore be regarded as an excitation above this vacuum, since taking $x_+ = 0$ indeed recovers the Minkowski spacetime.
Although the decoupled black hole solution \eqref{eq:4d-decoupled-BH} is asymptotically flat at leading order in the near-EVH decoupling limit, higher-order corrections --- following the framework of EMMD theory with a cosmological constant \cite{Lu:2013eoa} --- could in principle introduce an effective cosmological constant. 
However, such contributions do not capture the leading essential infrared physics of the near-EVH limit, and we will not pursue them further in this work.

This near-EVH limit of black holes can be further generalized to the non-extremal models, whose decoupled geometry can be conformally related to the non-extremal EMMD black holes. 
The corresponding near-EVH limit \eqref{eq:near-EVH-AdS6-cond-1} should be replaced by 
\begin{equation}\label{eq:near-EVH-AdS4-non-ext}
	q=b \sqrt{x_+ x_-} \epsilon + q^{(3)} \epsilon^3\,,
\end{equation}
which yields to the four dimensional decoupled black hole geometry 
\begin{equation}\label{eq:AdS6-decouple-nonSUSY}
	ds^2_{4} = - \frac{(x-x_+)(x-x_- ) d\tau^2}{(x+\frac{\sqrt{x_+x_-}}{b})^2} + \frac{x^2 dx^2}{(x-x_+) (x - x_-)}  + x^2 (d\theta_1^2 + \sin^2\theta_1 d\phi_1^2) \,.
\end{equation}
This geometry is conformal to the non-extremal EMMD black hole \eqref{eq:matching-EMMD-EVH} under the identification
\begin{equation}\label{eq:AdS6-nonextremal-ID}
	q_1+\mu =x_+,\quad  x_-=q_1, \quad  \frac{\sqrt{x_+ x_-}}{b} = q_2-q_1 , \quad x=\tilde{r} +q_1 \,,
\end{equation}
where the conformal factor can be absorbed into the dilaton, reflecting a change of frame.

\subsection{AdS$_7$ black hole} \label{sec:ADs7}

The AdS$_7$ black hole with two R-charges and three angular momenta has more fruitful parameter space. 
Such solutions are consistently reduced from $D=11$ supergravity with $S^4$ compact sphere \cite{Nastase:1999kf,Nastase:1999cb,Liu:1999ai}.
The bosonic part of the Lagrangian of the $D=7$ SO$(5)$ supergravity theory reads 
\begin{eqnarray} \label{eq:lagrangian-AdS7}
\mathcal{L}_7& =& R\star 1 + 2\left(
8 X_1 X_2 + \frac{4(X_1 +X_2)}{X_1^2 X_2^2} - \frac{1}{X_1^4 X_2^4}
\right) \star 1 - \frac{1}{2} \sum_{I=1}^2 d\varphi_I \wedge \star d\varphi_I \\ \nonumber
&& - \frac{1}{2} \sum_{I=1}^2 \frac{1}{X_I^2} F_{(2)}^I \wedge \star F_{(2)}^I -\frac{1}{2} X_1^2 X_2^2 F_{(4)} \wedge \star F_{(4)} + F_{(4)} \wedge A_{(3)} +F_{(2)}^1 \wedge F_{(2)}^2  \wedge A_{(3)} \,.
\end{eqnarray}
The most general AdS$_7$ black hole of this kind was worked out analytically in \cite{Bobev:2023bxl}, which includes all the known special solutions with one rotation turned off \cite{Wu:2011gp,Chow:2011fh}, equal charges \cite{Chow:2007ts} or equal angular momenta \cite{Chong:2004dy,Cvetic:2005zi}.
The most general solution to 7d gauged supergravity in the asymptotically static frame $(t,r,y,z,\phi_i)$ for $i=1,2,3$ is explicitly \cite{Bobev:2023bxl} 
\begin{align} \label{eq:AdS7-metric}
	\begin{split}
ds^2&= (H_1 H_2)^{\frac{1}{5}} \left[ 
- \frac{(1+r^2) (1-y^2)(1-z^2)}{\Xi_1 \Xi_2\Xi_3} dt^2 + \frac{(r^2+y^2)(r^2+z^2)}{U} dr^2
\right. \\
&+ \frac{(r^2+y^2)(y^2-z^2) y^2}{(1-y^2)(a_1^2-y^2)(a_2^2-y^2)(a_3^2-y^2)} dy^2 \\
&+ \frac{(r^2+z^2)(z^2-y^2) z^2}{(1-z^2)(a_1^2-z^2)(a_2^2-z^2)(a_3^2-z^2)} dz^2 \\
&+\frac{(r^2+a_1^2)(a_1^2-y^2)(a_1^2-z^2)}{\Xi_1(a_1^2-a_2^2)(a_1^2-a_3^2) } d\phi_1^2   
+\frac{(r^2+a_2^2)(a_2^2-y^2)(a_2^2-z^2)}{\Xi_2(a_2^2-a_1^2)(a_2^2-a_3^2) } d\phi_2^2 \\
& \left. +\frac{(r^2+a_3^2)(a_3^2-y^2)(a_3^2-z^2)}{\Xi_3(a_3^2-a_1^2)(a_3^2-a_2^2) } d\phi_3^2  + \frac{1-H_1^{-1}}{1-(s_2/s_1)^2 }K_1^2  + \frac{1-H_2^{-1}}{1-(s_1/s_2)^2} K_2^2
\right] \,.
	\end{split}
\end{align}
where for simplicity, we also take the cosmological constant $g^2=1$. 
The factors in the metric are defined as ($i,j,k=1,2,3$ are assumed to be different indices)
\begin{align}\label{eq:AdS7-def-function}
\begin{split}
& s_I= \sinh\delta_I, \quad c_I = \cosh \delta_I, \quad \Xi_i =1-a_i^2, \quad I=1,2 \\
& H_I= 1+ \frac{2m s_I^2}{(r^2+y^2)(r^2+z^2)} \\
U(r) &= \frac{(1+r^2)\prod_{i=1}^3 (r^2+a_i^2)}{r^2} -2m + m(s_1^2+s_2^2) \left(2r^2+\sum_{i=1}^3 a_i^2 \right) + \frac{4m^2 s_1^2 s_2^2}{r^2} \\
&- \frac{2m(s_1^2+s_2^2) a_1 a_2 a_3}{r^2} + \frac{2m(c_1-c_2)^2}{r^2}
(a_1+ a_2 a_3) (a_2+ a_1 a_3) (a_3+ a_1 a_2) \\
& + \frac{m(c_1-c_2)^2}{2} \Big[2\sum_{i=1}^3 a_i^2+8a_1 a_2 a_3 \\
&+(a_1+a_2+a_3) (a_2+a_3-a_1) (a_1+a_3-a_2) (a_1+a_3-a_2)\Big]  \\
K_1 &= \frac{c_1+c_2}{2s_1} \mathcal{A}[y^2,z^2,0] + \frac{c_1-c_2}{2s_1}\mathcal{Y}, \quad K_2 =\frac{c_1+c_2}{2s_2} \mathcal{A}[y^2,z^2,0] - \frac{c_1-c_2}{2s_2}\mathcal{Y} \\
&\mathcal{A}[y^2,z^2,0] =\frac{(1-y^2)(1-z^2)}{\Xi_1 \Xi_2 \Xi_3} dt -\sum_{i=1}^3 \frac{a_i(a_i^2-y^2) (a_i^2-z^2)}{\Xi_i(a_i^2-a_j^2)(a_i
	^2-a_k^2)} d\phi_i  \\
& \mathcal{Y} =  \frac{(1-y^2)(1-z^2) [1-(a_1^2+a_2^2+a_3^2)-2a_1 a_2 a_3]}{\Xi_1 \Xi_2 \Xi_3} dt \\
&\,\, + \sum_{i=1}^3 \frac{a_i (a_i^2-y^2) (a_i^2-z^2) [1-(a_i^2-a_j^2-a_k^2) + \frac{2a_j a_k}{a_i}]}{\Xi_i (a_i^2-a_j^2)(a_i^2-a_k^2)} d\phi_i \,,
\end{split}
\end{align}
and without loss of generality, the $y,z$ coordinates are taken in the region
\begin{equation}\label{eq:def-domain-yz}
	0 \le a_1 \le y \le a_2 \le z\le a_3 \le 1 \,.
\end{equation}

The parameterization in this metric differs slightly from that used in AdS$_5$ and AdS$_6$ black hole solutions, where redefining the charge parameter as $q_I = 2m s_I^2$ offers considerable convenience.
However, this simplification no longer applies to the AdS$_7$ solution given in \eqref{eq:AdS7-metric}, due to the presence of the $(c_1 - c_2)$ term in the blackening factor.
This expression becomes tractable only in the equal-charge case.
The event horizon is located at $r_+$, defined as the largest root of $U(r_+) = 0$.
In AdS$_{5,6}$ cases, one can invert this relation to express $m$ explicitly in terms of the horizon radius, leading to a convenient form for the blackening factor.
In contrast, for AdS$_7$, the function $U(r)$ is quadratic in $m$, which significantly complicates such an inversion.
This increased complexity enriches the structure of the parameter space and opens up new possibilities for exploring different EVH decoupling limits.

The thermodynamic quantities of AdS$_7$ black hole \eqref{eq:AdS7-metric}
are then computed by Komar integral \cite{Bobev:2023bxl}: 
\begin{eqnarray} \label{eq:AdS7theromo}
S &=& \frac{\pi^3}{4G_N \Xi_1 \Xi_2 \Xi_3} \frac{\sqrt{\mathcal{S}(r_+)}}{r_+} \,, \qquad T= \frac{r_+^2 U'(r_+)}{4\pi \sqrt{\mathcal{S}(r_+)}} \,, \\ \nonumber
J_1 &=&  \frac{\pi^2 m}{16G_N \Xi_1 \prod_{j=1}^3 \Pi_j}
\biggl[ 4a_1 c_1 c_2+ 4(1-c_1 c_2)(a_2 +a_1 a_3) (a_3+a_1 a_2) \\ \nonumber
&&+ (c_1-c_2)^2 \left( 
2a_2 a_3 + a_1 (1+2\Xi_1-\sum_{j=1}^3 \Xi_j)
\right) \left(1+2a_1 a_2a_3-\sum_{j=1}^{3} \Xi_j \right) \biggr]
\\ \nonumber
Q_1 &=& \frac{\pi^2 ms_1}{4G_N \Xi_1 \Xi_2 \Xi_3}  \Big[ 
2c_1 -(c_1 -c_2) (a_1^2+a_2^2+a_3^2+2a_1 a_2 a_3)
\Big] \,,
\end{eqnarray}
with the angular velocity given by 
\begin{eqnarray}
\Omega_1 &=& \frac{1}{\mathcal{S}(r_+)} \left[ 
\frac{1}{2} \left(  \prod_{i=1}^3 (r_+^2 +a_i^2) +2ms_1^2 (r_+^2-a_1 a_2 a_3)
\right)
\right. \\  \nonumber
&& \ \times  \Big(a_1(1+r_+^2)(r_+^2 +a_2^2)(r_+^2+a_3^2) +2ms_2^2 (a_1 r_+^2-a_2 a_3) \Big) \\ \nonumber
&&\ \  +\frac{1}{2} \left(  \prod_{i=1}^3 (r_+^2 +a_i^2) +2ms_2^2 (r_+^2-a_1 a_2 a_3)
\right)
\\  \nonumber
&& \ \times  \Big(a_1(1+r_+^2)(r_+^2 +a_2^2)(r_+^2+a_3^2) +2ms_1^2 (a_1 r_+^2-a_2 a_3) \Big) \\ \nonumber
&& - m(c_1-c_2)^2 (r_+^2 +a_2^2)(r_+^2+a_3^2)  \Big\{ 2(r_+^2-a_1)(a_1+a_2 a_3) (a_2+a_1 a_3) (a_3+a_1 a_2) \\ \nonumber
&&\left. + (1+a_1)^2 (a_1+ a_2 a_3)(1-a_1 -a_2 -a_3)(1-a_1+a_2+a_3) \Big\} \right] \,,
\end{eqnarray}
together with chemical potentials as 
\begin{eqnarray} 
\Phi_1&=& \frac{2m r_+^2 s_1 c_1 [\prod_{i}(r_+^2 +a_i^2) +2m s_2^2 (r_+^2 -a_1 a_2 a_3)]}{\mathcal{S}(r_+)} \\ \nonumber
&&- \frac{m r_+^2s_1(c_1 -c_2)}{\mathcal{S}(r_+)} \Big[ (a_1^2+a_2^2+a_3^2+2a_1 a_2 a_3) \left(
\prod_{i} (r_+^2+a_i^2) +2 ms_2^2 (r_+^2 -2a_1 a_2 a_3)
\right) \\ \nonumber
&& + 4ms_2^2 (a_1 +a_2 a_3) (a_2 +a_1 a_3) (a_3 +a_1 a_2)  \Big] \,,
\end{eqnarray}
where the function $\mathcal{S}$ is defined as
\begin{eqnarray}\label{eq:def-mathcalSfunction}
		\mathcal{S}(r) &=& \prod_{I=1}^2 \Big[(r^2+a_1^2)(r^2 +a_2^2)(r^2 +a_3^2)+2m s_I^2 (r^2-a_1 a_2 a_3) \Big]  \\ \nonumber
		&& + 2m(c_1 -c_2)^2 (r^2+a_1^2)(r^2 +a_2^2)(r^2 +a_3^2) (a_1+ a_2 a_3) (a_2+ a_1 a_3) (a_3+ a_1 a_2)\,.
\end{eqnarray}
The charges, angular momenta and chemical potentials with other indices can be worked out by permutating the indices. 
The energy is displayed as 
\begin{eqnarray} \nonumber
E &=& \frac{m\pi^2}{8G_N \Xi_1 \Xi_2 \Xi_3} \left[ \sum_{i=1}^3 \frac{2}{\Xi_i} -1 + \frac{5(s_1^2+s_2^2)}{2} + \frac{s_1^2+s_2^2}{2}
\sum_{i=1}^3 \left(
\frac{2(1+a_i^2-\Sigma_2-2\Pi_1)}{\Xi_i} -\Xi_i
\right)
\right] \\ \nonumber
&&+ \frac{m \pi (c_1 -c_2)^2}{32G_N \Xi_1 \Xi_2 \Xi_3} \Big[ 
-10\Sigma_2-16\Pi_1+11\Sigma_4 +13 \Pi_{22} +32 \Pi_1\Sigma_2 -3(\Sigma_6+5 \Pi_{42}+ 4\Pi_1^2) \\
&& \label{eq:AdS7-energy} 
-16 \Pi_1 \Sigma_2^2 +\Pi_{62} +3\Pi_{44} -5\Pi_1^2\Sigma_2 +8\Pi_1 (2\Pi_2 +\Pi_{42}) +\Pi_{1}^2 (\Sigma_4+3\Pi_{22}) \Big] \,,
\end{eqnarray}
where factors are defined as \cite{Bobev:2023bxl}
\begin{align}
	\begin{split}
\Sigma_n &= a_1^n+a_2^n+a_3^n\,, \qquad \Pi_n = a_1^na_2^na_3^n\,, \\
\Pi_{nm} &=a_1^n(a_2^m+a_3^m) + a_2^n(a_1^m+a_3^m) + a_3^n(a_1^m+a_2^m) \,.
	\end{split}
\end{align}
These quantities were checked in \cite{Bobev:2023bxl} to satisfy the first law of thermodynamics, as the consistency check of the validity of the solution. 
The thermodynamics of more special solutions were also studied in \cite{Larsen:2020lhg,Cassani:2019mms}.
The supersymmetry condition for this solution\footnote{Our conventions for the BPS condition follow \cite{Bobev:2023bxl}, differing from those for AdS$_{5,6}$ BPS black holes by the replacement $a_i \to -a_i$.} is given by
\begin{equation}\label{eq:AdS7-BPS}
	E = Q_1 +Q_2 -J_1 -J_2 -J_3 \,,
\end{equation}
which imposes a relation among the charges:
\begin{equation}\label{eq:SUSY-parameters-7}
	a_1 +a_2 +a_3 = \frac{2}{1-e^{\delta_1 +\delta_2}} \,.
\end{equation}
This is supplemented by a constraint on the horizon size which satisfies
\begin{equation}\label{eq:BPShorizon-AdS7}
	r_0^2 = \frac{a_1 a_2+a_2 a_3 +a_1 a_3 -a_1 a_2 a_3}{1-a_1 -a_2 -a_3}
\end{equation}
to fully determine the parameters describing the BPS black hole.
The chemical potentials in the BPS limit can be defined as 
\begin{equation}
	\Delta_I = \beta(\Phi_I+1),\qquad  \omega_i = \beta(\Omega_i +1) \,,
\end{equation}
which is subject to the constraint 
\begin{equation}
	\sum_{i=1}^3 \omega_i -2 \sum_{I=1}^2 \Delta_I =2\pi i \,.
\end{equation}

We are also interested in identifying possible (near)-EVH limits from these black hole solutions.
To systematically explore such limits, we adopt the following guidelines:
\begin{itemize}
	\item The near-horizon limit is defined as $\epsilon \to 0$, with the radial coordinate rescaled as $r = \epsilon x$ to render $\epsilon$ dimensionless.
	\item The entropy $S$ and temperature $T$ are assumed to scale with $\epsilon$ as
	\begin{equation}\label{eq:mn-nearEVH}
		S \sim \epsilon^n, \qquad \quad T \sim \epsilon^{m-n+1}, \qquad m,n\in \mathbb{Z} \,,
	\end{equation}
which in turn fixes the scaling of the entropy function and the blackening factor:
\begin{equation}\label{eq:scalingSSU}
	\mathcal{S}(r_+) \sim \epsilon^{2n+2}, \qquad U'(r_+)\sim\epsilon^m \,.
\end{equation}
For simplicity, we restrict our analysis to integer values of $m$ and $n$.
\end{itemize}
Our tasks are to find the scaling of $a_i,q_I$ parameters to match the possible \eqref{eq:scalingSSU}, in order to achieve the near-EVH scalings of entropy and temperature \eqref{eq:mn-nearEVH}. 
In this paper, we will only discuss $S\sim T^k$ for some integer $k$, which could be potentially interpretable as field theory with scaling symmetry but excluding Lifshitz scaling \cite{Kachru:2008yh} or hyperscaling violation scalings \cite{Dong:2012se}. 

Therefore, the possible solutions are 
\begin{align}\label{eq:EVH-AdS7-condi}
	\begin{split}
(m,n)=(1,1): & \qquad S\sim T \\
(m,n)=(2,2): & \qquad S\sim T^2 \\
(m,n)=(3,3): & \qquad S\sim T^3 \,.
	\end{split}
\end{align}
Within these choices, the $S\sim T$ near-EVH limits are well-known to produce the pinched BTZ-like geomtries in the IR.  
Just as the near-EVH BTZ black holes emergent in static R-charged AdS$_5$ black holes are supported by two order $G_N^{-1}$ charges and one perturbative R-charge \cite{Balasubramanian:2007bs}, the BTZ in the near-EVH of AdS$_7$ black holes can be supported by two angular momenta of order $G_N^{-1}$ and one perturbative angular momentum. 

More nontrivial are the scalings $S \sim T^2$ and $S \sim T^3$. Previous analysis of the near-EVH limit in AdS$_6$ identifies $S \sim T^2$ as signaling the emergence of a $D=4$ EMMD black hole in the infrared. In AdS$_7$, however, no suitable parameter regime exists that realizes a consistent near-EVH limit with $S \sim T^2$. On the other hand, the case $S \sim T^3$ has previously been identified in the near-EVH limit of AdS$_7$ black holes \cite{Goldstein:2019gpz}. In the following, we present the specific parameter choices that define each of these limits and describe the corresponding infrared geometries.

\subsubsection*{Example: $D=3$ decoupled geometry}

It is observed \cite{Chen:2006xh,David:2020ems} that AdS$_3$ and the BTZ geometry can appear in the near-horizon limits of the AdS$_7$ black hole. 
Such limits can be achieved by taking two non-vanishing angular momenta and treating the third angular momentum to be perturbative.

We can simply take the near-EVH limit as
\begin{align}
	\begin{split}
s_I\equiv 0, \qquad a_3=r_+ = 0, \qquad r= \epsilon x, 
	\end{split}
\end{align}
while the near-EVH limit is defined as 
\begin{equation}\label{eq:near-EVH-AdS7-BTZ}
a_3= \lambda \epsilon^2, \qquad r_+ = \epsilon x_+ \,.
\end{equation}
Notice these exactly reduce to the Kerr-AdS$_7$ black holes  \cite{Chen:2006xh,Gibbons:2004js} and are inconsistent with supersymmetry condition \eqref{eq:SUSY-parameters-7}.
Their thermodynamic quantities are 
\begin{align}
	\begin{split}
E & = \frac{m \pi^2}{4G_N\Xi_1 \Xi_2 \Xi_3} \left(\sum_{i=1}^3 \frac{1}{\Xi_i} - \frac{1}{2}\right), \qquad  J_i = \frac{m a_i \pi^2}{4G_N\Xi_i(\prod_{j=1}^3 \Xi_j)},  \\
 S&= \frac{\pi^2}{4G_N r_+} \prod_{i=1}^3 \frac{r_+^2+a_i^2}{\Xi_i}, \qquad \Omega_i = \frac{a_i \Xi_i}{r_+^2+a_i^2} \,,\\
T & = \frac{1}{2\pi} \left[r_+ (1+r_+^2)\sum_{i=1}^3 \frac{1}{r_+^2+a_i^2} - \frac{1}{r_+} \right] \,.
	\end{split}
\end{align}
The near-EVH limits defined in \eqref{eq:near-EVH-AdS7-BTZ} thus results in 
\begin{equation}
	S\sim T \sim\epsilon, \qquad \lim_{\epsilon \to 0}\frac{S}{T} = \frac{\pi^3a_1^2 a_2^2}{2 G_N(1-a_1^2)(1-a_2^2)(1+\frac{1}{a_1^2} + \frac{1}{a_2^2} - \frac{\lambda}{r_+^4})} \,,
\end{equation}
which confirms this is a reasonable near-EVH limit. 
The black hole geometry in the limit \eqref{eq:near-EVH-AdS7-BTZ} is of the following ansatz
\begin{equation}
	\begin{aligned}
		&ds_7^2= \epsilon^2 \frac{y^2z^2}{a_1^2a_2^2} ds_3^2+\frac{y^2(y^2-z^2)}{(1-y^2)(a_1^2-y^2)(a_2^2-y^2)}dy^2
		+\frac{z^2(z^2-y^2)}{(1-z^2)(a_1^2-z^2)(a_2^2-z^2)}dz^2\\
		+&\frac{(1-y^2)(a_1^2-y^2)(a_2^2-y^2)}{y^2(y^2-z^2)}\left(-\frac{a_1(a_1^2-z^2)}{\Xi_1(a_1^2-a_2^2)}d\phi_1-\frac{a_2(a_2^2-z^2)}{\Xi_2(a_2^2-a_1^2)}d\phi_2\right)^2\\
		+&\frac{(1-z^2)(a_1^2-z^2)(a_2^2-z^2)}{z^2(z^2-y^2)}\left(-\frac{a_1(a_1^2-y^2)}{\Xi_1(a_1^2-a_2^2)}d\phi_1-\frac{a_2(a_2^2-y^2)}{\Xi_2(a_2^2-a_1^2)}d\phi_2\right)^2 \,,
	\end{aligned}
\end{equation}
where the three dimensional part is a pinched AdS$_3$ geometry in the EVH limit
\begin{equation}\label{eq:near-EVHAdS3-AdS7}
	ds_3^2 = - \frac{ x^2 }{l_3^2}{d}\tau^2+\frac{l_3^2}{ x^2 }dx^2+  x^2d\tilde{\phi}_3^2, \qquad  l_3^2= \frac{a_1^2a_2^2}{a_1^2+a_2^2+a_1^2a_2^2}\,,
\end{equation}
and a pinched extremal BTZ in the near-EVH limits with extremality condition $r_+ =r_- $: 
\begin{equation}
	ds_3^2 = -\frac{ (x^2-x_+^2)^2}{l_3^2x^2}d\tau^2+\frac{l_3^2x^2}{(x^2-x_+^2)^2}dx^2+ x^2\bigg(\frac{x_+^2}{l_3x^2}d\tau-d\tilde{\phi}_3\bigg)^2\,.
\end{equation}

In string theory, the local AdS$_3$ geometry arises as the near-horizon limit of a bound state consisting of $n_1$ D1-branes, $n_5$ D5-branes, and a momentum $P$ along a common compact direction. 
The D1-branes, viewed as instantons within the D5-branes, combine with the momentum charge to produce a two-dimensional conformal field theory known as the D1–D5 CFT. 
This theory has central charge $c = 6 n_1 n_5$, which counts the number of degrees of freedom in the bound state. 
We refer to \cite{David:1999dx} for a comprehensive review.
In the context of R-charged AdS$_5$ black holes, a local AdS$_3$ geometry similarly emerges in the near-EVH limit when two R-charges are present. 
The three R-charges, corresponding to the Cartan generators of the SO(6) R-symmetry, represent different species of giant gravitons. The effective theory arising from two species of giant gravitons—localized at their intersection—has degrees of freedom proportional to the product of the two macroscopic R-charges, yielding a central charge $c \sim Q_1 Q_2$ \cite{Balasubramanian:2007bs,Fareghbal:2008ar}.
Similarly, for Kerr–AdS$_7$ black holes in the near-EVH limit, the central charge of the dual CFT$_2$ associated with the emergent AdS$_3$ is proportional to the product of two angular momenta:
\begin{equation}
c\sim 	l_3 \sim a_1 a_2 \sim J_1 J_2 \,.
\end{equation}
This suggests that the CFT$_2$ dual to the emergent AdS$_3$ resides on the intersection of dual giant-like objects in AdS$_7$ \cite{Gupta:2023vwn}.

\subsubsection*{Example: $D=5$ decoupled geometry}

As the $S\sim T^3$ near-EVH limit exists in equal-charge AdS black holes with three distinct angular momenta, we can just take the equal-charge solution \cite{Chow:2007ts} to examine its near-EVH geometry for simplicity. 
Therefore, we take $q=2m s_1^2 = 2ms_2^2$. 
Especially in the BPS limit, the horizon size is completely fixed by rotation parameters \eqref{eq:BPShorizon-AdS7}.
Therefore, the EVH-BPS limit related to taking $r_+ \sim \epsilon$ is completely fixed by the scalings of rotation parameter,  independent of whether charges are taken equal or not. 

The EVH limit to approach $S\sim T^3$ is defined as 
\begin{equation}
a_1 =a_2 =0, \quad q=0, \quad r=\epsilon x, \quad x_+ =0, \quad t= \epsilon \tau \,.
\end{equation}
In order to have well-defined geometry in this limit and consistent with \eqref{eq:def-domain-yz}, the parametrization
\begin{equation}
	y^2 = a_1^2 \cos^2\theta_1 + a_2^2 \sin^2\theta_1, \qquad z= a_3 \cos \theta_2 \,,
\end{equation}
automatically translate from the Jacobi coordinate \eqref{eq:Jacobi-transformation} to spherical coordinate $(\theta_1,\theta_2)$.
In this coordinate, the EVH geometry of AdS$_7$ black hole \eqref{eq:AdS7-metric} is paramterized by coordinates $(t,r,\theta_1,\theta_2,\phi_i)$ for $i=1,2,3$:
\begin{align}\label{eq:EVH-geometry-AdS7}
	\begin{split}
		ds^2 &= - \frac{(1-a_3^2 \cos^2\theta_2)}{1-a_3^2} (1+r^2) dt^2 + \frac{r^2+a_3^2 \cos^2\theta_2}{(r^2+1)(r^2+a_3^2)} dr^2 \\
		&+ r^2\cos^2\theta_2( d\theta_1^2 + \sin^2\theta_1 d\phi_1^2 + \cos^2\theta_1 d\phi_2^2) \\
		& + \frac{r^2+a_3^2 \cos^2\theta_2}{1-a_3^2 \cos^2\theta_2} d\theta_2^2 + \frac{r^2+a_3^2}{1-a_3^2} \sin^2\theta_2 d\phi_3^2 \,.
	\end{split}
\end{align}
At the location $\theta_2 \to 0$, the geometry decouples to global AdS$_5$ spanned by $(t,r,\theta_1,\phi_{1,2})$ and $S^2$ spanned by $(\theta_2,\phi_3)$. 
We take the metric ansatz of \cite{Chow:2007ts} and implement the following near-EVH limits
\begin{equation}\label{eq:near-EVHAdS7-T5}
	a_1 = \lambda_1 \epsilon^2, \quad a_2= \lambda_2 \epsilon^2, \quad q=a_3 x_+^2 \epsilon^2 + q^{(4)}\epsilon^4, \quad r= \epsilon x + y_4 \epsilon^3  \,.
\end{equation}
It is convenient to define the new coordinate $y=w \epsilon^2$, which satisfies 
\begin{equation}\label{eq:redef-coodinate-w}
	w^2 = \lambda_1^2 \cos^2\theta_1 +\lambda_2^2\sin^2\theta_1 \,.
\end{equation}
The entropy and temperature in the near-EVH limit \eqref{eq:near-EVHAdS7-T5} are then fixed by the perturbative ansatz \eqref{eq:near-EVHAdS7-T5}
\begin{align}\label{eq:AdS7STnearEVHT5}
	\begin{split}
		S&= \frac{\pi^3 a_3 x_+^3 \epsilon^3}{4G_N(1-a_3)} +\mathcal{O}(\epsilon^4) \\
		T&= \frac{(1+a_3+a_3^2)x_+^4 + a_3^2 \lambda_1 \lambda_2 - q^{(4)} a_3 + 2y_4 a_3^2x_+}{a_3(1+a_3) \pi x_+^3} \epsilon + \mathcal{O}(\epsilon^2) \,,
	\end{split}
\end{align}
confirming the existence of $S\sim T^3$ scaling in this limit for generic values of $q^{(4)}$. 
The special value of $q^{(4)}$ making the temperature vanishing is determined as 
\begin{equation}
	q^{(4)}=\left(1+\frac{1}{a_3} +a_3\right) x_+^4+a_3\lambda_1\lambda_2+2y_4a_3x_+ \,.
\end{equation}
In this paper, we will not go to this fine tuning values as we are interested in the general near-EVH decoupling limits. 
The relations \eqref{eq:AdS7STnearEVHT5} are valid even without imposing the BPS condition, thus generalizing the EVH-BPS limit of AdS$_7$ proposed in \cite{Goldstein:2019gpz}.
If the supersymmetry condition is imposed, the horizon size satisfies $x_+ =x_0$, where 
\begin{equation}\label{eq:AdS7-EVH-BPS-horizon}
	x_0^2 = \frac{a_3(\lambda_1+\lambda_2)}{1-a_3}\,,
\end{equation}
and the entropy of the EVH-BPS black hole is fixed as
\begin{equation}\label{eq:BPS-AdS7entropy-EVH}
	S= \frac{\pi^3}{4G_N} \left(\frac{a_3}{1-a_3} \right)^{\frac{5}{2}} (\lambda_1+\lambda_2)^{\frac{3}{2}}\epsilon^3 \,.
\end{equation}

We now implement the near-EVH limit given in \eqref{eq:near-EVHAdS7-T5} for the AdS$_7$ black hole metric. As in the AdS$_6$ case, the angular coordinate $\theta_2$, which ranges between $0$ and $\frac{\pi}{2}$, governs the emergence of distinct decoupled geometries in the near-horizon region.
In the limit $\theta_2 \to 0$, the metric decouples into the following five-dimensional form:
\begin{equation}
ds_7^2 =H^{\frac{2}{5}}\left[ \epsilon^2 ds_5^2  + \frac{a_3^2 }{(1-a_3^2)} d\Omega_2^2 \right], \qquad d\Omega_2^2  = d\theta_2^2 + \sin^2\theta_2 d\phi_3^2 \,,
\end{equation}
where the five-dimensional geometry is given by
\begin{align}\label{eq:EMMD-D5-extre}
	\begin{split}
ds_5^2 &= - \frac{(x^2-x_+^2)^2}{(x^2+ \frac{x_+^2}{a_3})^2} d\tau^2 +\frac{x^4 dx^2}{(x^2-x_+^2)^2} + x^2  d\Omega_3^2 \,, \\
d\Omega_3^2 & = d\theta_1^2 +\sin^2\theta_1 d\phi_1^2 +\cos^2\theta_1 d\phi_2^2 \,.
\end{split}
\end{align}
Moreover, the geometry in \eqref{eq:EMMD-D5-extre} is conformally related to a $D=5$ EMMD black hole with exponents $(N_1, N_2) = (1, 2)$. In the extremal case, the latter takes the form (cf. \eqref{eq:EMMD-5d}):
\begin{equation}
d\tilde{s}_5^2 = \frac{(x^2+q_2-q_1)^{\frac{2}{3}}}{x^{\frac{4}{3}}} \left[-\frac{(x^2-q_1)^2 dt^2}{(x^2+q_2-q_1)^2} + \frac{x^4 dx^2}{(x^2-q_1)^2 } + x^2 d\Omega_3^2
		\right] \,.
\end{equation}
By identifying the EMMD charges $q_1$ and $q_2$ (assuming $q_1 < q_2$ without loss of generality) with parameters of the AdS$_7$ solution, we obtain:
\begin{equation}
	x_+^2 = q_1, \qquad \frac{x_+^2}{a_3} =q_2 -q_1, \qquad q_2 >q_1  \,.
\end{equation}

The near-EVH limit of AdS$_7$ black holes exhibits several structural similarities with the AdS$_6$ case analyzed in Section \ref{sec:ADs6}.
First, in both instances, the emergent decoupled near-EVH geometry for an AdS$_D$ black hole is an asymptotically flat EMMD-type solution in $(D-2)$ dimensions. 
In the formal limit where the Boyer–Linquist rotation parameter tends to infinity, these geometries reduce to a Reissner–Nordström black hole, a regime that lies outside the domain of validity of the near-EVH approximation.
Second, analogous to the AdS$_6$ result, the correspondence between the near-EVH limit of the AdS$_7$ black hole and the five-dimensional EMMD description also extends to non-extremal configurations. 
This generalization is achieved by modifying the near-EVH scaling in \eqref{eq:near-EVHAdS7-T5} to incorporate a subleading correction:
\begin{equation}
	q= a_3 x_+ x_- \epsilon^2+ q^{(4)} \epsilon^4 \,,
\end{equation}
which yields the following decoupled five-dimensional metric conformal to the EMMD black hole in $D=5$:
\begin{equation}
	ds^2_5 = - \frac{(x^2-x_+^2)(x^2-x_-^2)}{(x^2+ \frac{x_+ x_-}{a_3})^2} d\tau^2 +\frac{x^4 dx^2}{(x^2-x_+^2)(x^2-x_-^2)} + x^2  d\Omega_3^2 \,.
\end{equation} 
Finally, in both AdS$_6$ and AdS$_7$ settings, the EMMD decoupling limit becomes fully explicit only when the direction cosine coordinate is set to a special value, $\theta_2 \to 0$. 
This choice forces the coordinate $z$ to approach its upper bound, thereby reducing the $S^{2}$ sphere (spanned by $\mu_i^2$) to an $S^1$. The potential physical implications of this parameter restriction will be addressed in Section \ref{ssec:discussion-D-67}.

\subsection{Discussion}
\label{ssec:discussion-D-67}

The AdS$_{D}$ black holes in $D=6,7$ analyzed in this section exhibit an infrared scaling relation $S \sim T^{D-4}$, reflecting the emergence of an effective $(D-2)$-dimensional geometry in the decoupling limit. 
These on-shell geometries are structurally conformal to EMMD black holes \cite{Lu:2013eoa}.
Furthermore, the rich parameter space of AdS$_7$ allows three-dimensional BTZ geometries to emerge in the decoupling limit as well. 
The decoupled geometries discussed in this work can be viewed as specific ground states—analogous to AdS$_2$—that encode the ground-state degeneracy of the higher-dimensional black holes. A key distinction, however, is that the near-EVH limit permits excitations along additional directions, giving rise to a lower-dimensional effective field theory. 
In this picture, extremal EMMD black holes—characterized by vanishing temperature—correspond to the ground states of these lower-dimensional effective theories.
A structural distinction also exists between the AdS$_3$-type near-EVH geometry and the EMMD-type geometry. The former corresponds to a pinched, locally AdS$_3$ spacetime, in which the compact $S^1$ direction has period $2\pi \epsilon$ and the $dr^2$ component remains of order $\epsilon^0$ \cite{deBoer:2010ac,Johnstone:2013eg,Goldstein:2019gpz}. 
In contrast, the near-EVH limits of AdS$_{D=6,7}$ exhibit a decoupling structure of the form
\begin{equation}
	ds_{D}^2  = \epsilon^2 ds^2_{\text{EMMD}}+ d\Omega^2_2 \,,
\end{equation}
which resembles a class of non-relativistic geometries known as M$p$T spacetimes \cite{Blair:2023noj}. We will revisit these observations in Section \ref{sec:discussion}.

In both the near-EVH limits of AdS$_6$ and AdS$_7$
black holes, the lower-dimensional EMMD geometry becomes explicit when the limit $\theta_2 \to 0$, , which effectively reduces the  $S^2$ of the direction cosine to an $S^1$. 
Given the analogous structures in the two models, we shall use the AdS$_6$ to illustrate the underlying physical picture.
From a string-theory viewpoint, the AdS$_6$
background is supported by a D4–D8 bound state in massive type‑IIA supergravity \cite{Cvetic:1999un}.
The limit $\theta_2 \to 0$ selects a sector in which excitations are confined to modes that do not probe the shrinking direction—i.e., collective low‑energy modes of the bound state that are effectively four‑dimensional.
The resulting EMMD theory—with its metric, two gauge fields, and a dilaton—thus provides a consistent infrared description of this non‑conformal, decoupled sector. Its microscopic origin should be traceable to an appropriate subsector of the dual five‑dimensional SCFT.
Because a more detailed understanding of this system in terms of giant gravitons is still lacking, we leave a more complete analysis for future work.

Several limitations in our study deserve mention.
First, we have restricted our analysis to (near)-EVH limits in which the rotation parameters scale with $\epsilon$ as integer powers, following the approach of \cite{Johnstone:2013eg,Goldstein:2019gpz}.
While most generic scaling choices do not yield a well-defined decoupling limit with a smooth horizon geometry, it may be worthwhile to explore more exotic scalings—such as those involving non-integer powers—which could be dual to other decoupled sectors in the holographic field theory.
Second, our discussion has been confined to on-shell solutions of supergravity in $D=6,7$ \cite{Chow:2007ts,Chow:2008ip}.
An interesting open question is whether these supergravity Lagrangians \eqref{eq:lagrangian-AdS6} and \eqref{eq:lagrangian-AdS7} can be related off-shell to the EMMD theory Lagrangian, for example \eqref{eq:EMMD-action}.
Such a connection would help clarify how EMMD theories emerge from fine-tuned decoupled vacua of ten-dimensional string theory or 11D supergravity via specific brane configurations.

As reviewed in Appendix \ref{appendix:EMMD}, the supergravity framework admits three distinct classes of EMMD black hole geometries, labeled by a pair of integers $(N_1, N_2)$ satisfying
\begin{equation}\label{eq:integerN1N2constraint}
	N_1 + N_2 = \frac{2(D-2)}{D-3}.
\end{equation}
For $D > 5$, this constraint cannot be satisfied by integer pairs—a fact that may be related to the property that AdS$_7$ black hole is the highest-dimensional AdS black hole embeddable in string theory with a known superconformal field theory dual.
For lower dimensions, the possible integer pairs (taking $N_1 < N_2$ without loss of generality due to symmetry) are:
\begin{align}
	\begin{split}
		D=4: &\qquad (N_1,N_2) = (2,2), \quad (1,3), \\
		D=5: &\qquad (N_1,N_2) = (1,2).
	\end{split}
\end{align}
It remains unclear why, although the near-EVH limits of AdS$_{6,7}$ black holes yield EMMD black holes of types $(2,2)$ and $(1,2)$ respectively, while the $(1,3)$ type—to the best of our knowledge—does not appear in the near-EVH limits studied here.
It is also possible that such a solution may emerge from the near-EVH limit of as-yet-unknown supergravity solutions in $D=6$ or $7$, which would be an interesting direction for future work.

\section{Holography with $S\sim T^{D-4}$}\label{sec:holography}

The entropy of BPS AdS$_D$ black holes is known from microstate counting in the dual superconformal field theory \cite{Choi:2018fdc,Choi:2019miv,Nahmgoong:2019hko,Bobev:2025xan,Hosseini:2018dob, Crichigno:2020ouj}.
These results can be applied to study the entropy of the emergent $(D-2)$-dimensional geometries arising in the near-extremal vanishing horizon (EVH) limit.
Although the extremization of the entropy functional in $D=6$ and $7$ exhibits certain similarities, it differs from the corresponding extremization in the near-EVH limit of AdS$_5$.
This further confirms that the near-EVH decoupling geometry in $D=6,7$, with entropy scaling as $S \sim T^{D-4}$, is not of BTZ type.

Consider the AdS$_6$ black hole as an example. Its entropy is given by the extremization of the functional \cite{Choi:2018fdc,Choi:2019miv}
\begin{align}\label{eq:entropyfunctional-AdS6}
	\begin{split}
S &= i \frac{\pi}{3 G_N} \frac{\Delta^3}{\omega_a \omega_b} + \Delta Q +\omega_a J_a +\omega_b J_b + \Lambda (3\Delta-\omega_a-\omega_b+2\pi i) \,,
	\end{split}
\end{align}
where $\Lambda$ is a Lagrange multiplier enforcing the constraint among chemical potentials.
Extremizing this functional via the saddle-point approximation,
\begin{equation}\label{eq:saddle-point-AdS6}
\frac{\partial S}{\partial \Delta} = \frac{\partial S}{\partial \omega_a}  = \frac{\partial S}{\partial \omega_b} =0\,,
\end{equation}
simplifies the entropy expression to
\begin{equation}\label{eq:entropy-Mulitpler}
	S= 2\pi i \Lambda \,.
	\end{equation}
For the entropy to be real with real physical charges, $\Lambda$ must be purely imaginary. 
This leads to two constraint equations involving $S$ \cite{Choi:2018fdc}:
\begin{align}\label{eq:AdS6-SQJconstraint}
	\begin{split}
& Q \frac{S^2}{4\pi^2} +  \frac{J_a +J_b}{6G_N} S- \frac{Q^3}{27} =0 \\
& \left( \frac{S}{2\pi}\right)^3- \frac{\pi}{3G_N} \left( \frac{S}{2\pi}\right)^2 - \frac{Q^2}{6\pi} S + \frac{\pi}{3G_N} J_a J_b=0 \,.
	\end{split}
\end{align}
Solving these gives the BPS entropy in terms of the charges and angular momenta:
\begin{equation}
	S= \frac{\pi}{Q} \left[ 
  \sqrt{(J_a+J_b)^2 \frac{\pi^2}{9G_N^2} + \frac{4}{27} Q^3}	 -(J_a+J_b) \frac{\pi}{3G_N}
	\right] \,,
	\end{equation}
subject to the charge constraint:
\begin{align}
	\begin{split}
& \frac{1}{Q} \left[ 
\sqrt{(J_a+J_b)^2 \frac{\pi^2}{9G_N^2} + \frac{4}{27} Q^3}	 -(J_a+J_b) \frac{\pi}{3G_N}
\right] \\
=& \frac{2\pi}{3G_N }Q^2 \frac{Q^2 +(J_a +J_b) Q -27 J_a J_b}{3(J_a+J_b)^2 \frac{\pi^2}{G_N^2} + 3\frac{\pi^2}{G_N^2} (J_a+J_b) Q - 8Q^4} \,.
	\end{split}
\end{align}
In the EVH limit $Q = J_a = 0$, the right-hand side vanishes identically, while the left-hand side remains proportional to the entropy $S$.
Thus, the entropy also vanishes in this limit.

In the near-EVH limit given by \eqref{eq:near-EVH-AdS6-cond-1} together with the BPS condition, the entropy expression \eqref{eq:near-EVH-bpsentropy} can also be reproduced by extremizing the functional in \eqref{eq:entropyfunctional-AdS6}.
This mechanism differs significantly from that of the near-EVH entropy in AdS$_5$ (see \eqref{eq:AdS5-S-lnZ}, where the entropy scales as $N^2\epsilon$ due to an extremization between terms of order $N^2$ and $N^2\epsilon^2$, leading to the emergence of a 2D Cardy formula \cite{Goldstein:2019gpz}. 
In the AdS$_6$ case, the thermodynamic quantities scale as follows:
\begin{align}\label{eq:near-EVH-thermo}
	\begin{split}
& S\sim	i \frac{\pi}{3G_N} \frac{\Delta^3}{\omega_a \omega_b}  \sim N^{\frac{5}{2}} \epsilon^2,\qquad \Delta\sim \omega_b \sim \epsilon, \quad Q\sim J_b\sim N^{\frac{5}{2}} \epsilon \\
& J_a \sim N^{\frac{5}{2}} \epsilon^3, \qquad \omega_a \sim 2\pi i +\mathcal{O}(\epsilon) \,.
	\end{split}
\end{align}
Since the angular momentum $J_a$ is subleading compared to $J_b$ and $Q$, it can be neglected at leading order. 
Incorporating the near-EVH scaling relations \eqref{eq:near-EVH-thermo} with the entropy functional and the chemical potential constraint can be shown to correctly reproduce the black hole entropy in the near-EVH limit \eqref{eq:near-EVH-bpsentropy}.

The entropy of AdS$_7$ black hole is reproduced by following functional \cite{Bobev:2023bxl, Zaffaroni:2019dhb,Choi:2018hmj,Cassani:2019mms}
\begin{align}
	\begin{split}
S &= - \frac{\pi^2}{8G_N} \frac{\Delta^4}{\omega_{1} \omega_2 \omega_3} + Q \Delta - \sum_{i=1}^3 J_i\omega_i - \Lambda \left(\sum_{i=1}^3 \omega_i -4\Delta-2\pi i \right)\,.
	\end{split}
\end{align} 
This can be derived from either 6d SCFT or the 5d $\mathcal{N}=2$ Yang-Mills theory on $S^5$ \cite{Kantor:2019lfo}.
The entropy evaluated as saddle point approximation is still captured by the Lagrange multipler \eqref{eq:entropy-Mulitpler} due to the homogeneity of this functional. 
To have real values of physical entropy with real charges and angular momenta, the entropy are determined by two equations 
\begin{align}\label{eq:BPSentropy-constraint-AdS7}
	\begin{split}
  &S^{2}=\frac{4\pi^{4}(J_{1}J_{2}+J_{2}J_{3}+J_{1}J_{3})-2G_N\pi^2Q^{3}}{\pi^{2}-8G_NQ}\\
&\frac{2S^{4}}{\pi^{4}}+\left(\frac{J_{1}+J_{2}+J_{3}}{G_{N}}-\frac{3Q^{2}}{\pi^{2}}\right)S^{2}+ \frac{Q^{4}}{8}-\frac{4\pi^{2}}{G_{N}}J_{1}J_{2}J_{3}=0  \,.
	\end{split}
\end{align}
In the exact EVH limit where $J_{1,2}=Q=0$, the constraint \eqref{eq:BPSentropy-constraint-AdS7} indicates the entropy vanishing, consistent with the computation on gravity side.
In the near-EVH limit \eqref{eq:near-EVHAdS7-T5}, the thermodynamic quantities scale with $\epsilon$ as 
\begin{equation}
\omega_1 \sim \omega_2 \sim i \pi, \qquad \Delta\sim \omega_3\sim \epsilon, \qquad J_1\sim J_2\sim N^3 \epsilon^4, \qquad J_3 \sim Q \sim N^3\epsilon^2 \,.
\end{equation}
Therefore, the contributions from angular momenta $J_{1,2}$ are completely subleading in the near-EVH limit as $N^3\epsilon^3$ is fixed. 
This phenomenon is similar to the extremization of AdS$_6$ entropy functional instead of AdS$_5$, where the subleading angular momenta are crucial to derive the Cardy formula. 

A few remarks are in order. 
\begin{itemize}
\item First, the contribution from the close to vanishing angular momenta is subleading, resulting in an inhomogeneous entropy functional that does not separate into a two-step extremization procedure—consistent with the fact that the near-EVH geometry is not of BTZ type. 
\item Second, we propose that the entropy functional in the limit \eqref{eq:near-EVH-thermo} should be interpreted as that of an EMMD black hole. It is essential to note that EMMD black holes do not carry angular momenta or support angular velocities \cite{Lu:2013eoa}. To establish this correspondence, the quantities in \eqref{eq:near-EVH-thermo} must therefore be mapped to the thermodynamic variables of EMMD black holes in the decoupled limit.
Furthermore, the explicit form of the EMMD metric exhibits a mild dependence on the coordinate $\theta_2$, which is integrated out in the near-EVH black hole entropy expressions \eqref{eq:near-EVH-bpsentropy} and \eqref{eq:BPS-AdS7entropy-EVH}. As a result, the near-EVH limit of the black hole entropy effectively averages over a family of EMMD geometries, making the matching between the EMMD black hole entropy and the near-EVH limit a more subtle task. This additional degree of freedom may be absorbed into a redefinition of the lower-dimensional Newton constant.
\item Finally, the EVH limits are defined to have certain subset of charges and angular momenta vanishing, which correspond to the enhanced supersymmetry. 
These are equivalent to define appropriate subsectors of the superconformal field theories dual to the AdS$_{6,7}$ black holes. 

\end{itemize}
All these holographic aspects, however, lie beyond the scope of the present work.

\section{Conclusion and future work}\label{sec:discussion}

In this work, we systematically explore possible near-horizon decoupled geometries in higher-dimensional AdS black holes embedded in string theory and M-theory \cite{Duff:1999rk}.
Thanks to their high dimensionality and rich parameter space, the $D=6,7$ black holes support model-dependent  decoupled geometries of dimension $D \geq 4$ that are conformal to black hole solutions of EMMD gravity.  
These generalize the universal AdS$_2$ throats appearing in the near-horizon limits of extremal black holes. 
(Note that our analysis is restricted to the leading order of the near-EVH limits, which might be insensitive to the presence of a cosmological constant in the decoupled EMMD geometries.)

As summarized in Table \ref{table:summaryre}, the emergence of these distinct decoupled geometries carries several crucial holographic implications. Most notably, the fact that the near-EVH limits in $D=6,7$ yield EMMD spacetimes rather than pure AdS$_{D-2}$ factors directly indicates that the decoupled low-energy sectors do not exhibit standard conformal invariance. Instead, the presence of a running dilaton in the EMMD solutions suggests that the dual infrared effective field theories are potentially governed by generalized scaling symmetries or hyperscaling-violating properties \cite{Ogawa:2011bz,Huijse:2011ef,Dong:2012se}, rather than being traditional CFTs.
\begin{table}[H]
	\centering
\begin{tabular}{|c|c|}
	\hline
Black hole	& Allowed near horizon decoupled geometry \\
	\hline
AdS$_4$	&AdS$_2$, BTZ  \\
	\hline
AdS$_5$	&AdS$_2$,  BTZ \\
	\hline
AdS$_6$	&AdS$_2$, EMMD in $D=4$ \\
	\hline
AdS$_7$	&AdS$_2$, BTZ or EMMD in $D=5$   \\
	\hline
\end{tabular}
\caption{The summary of the decoupled geometries appear in the near-EVH limits. }
	\label{table:summaryre}
\end{table}

The first open question is to understand the microscopic states dual to the classes of decoupled EMMD black holes \eqref{eq:4d-decoupled-BH} and \eqref{eq:EMMD-D5-extre}, which are asymptotically flat \cite{Lu:2013eoa,Lu:2025eub}. By rigorously embedding these non-AdS geometries into the well-defined near-EVH limits of AdS$_6$ and AdS$_7$, we establish a controlled, top-down holographic framework. This provides a clear pathway to tackle the microstate counting problem for non-AdS black holes. Specifically, the distinct thermodynamic scaling $S \sim T^{D-4}$ observed in these limits serves as a macroscopic prediction for the density of states, setting a concrete and precise target for future field-theoretic computations and the evaluation of partition functions in the dual IR theory. The field theories dual to these black holes should be viewed as decoupled subsectors in the higher-dimensional dual SCFTs in $D=5,6$ \cite{Bobev:2025xan,Nahmgoong:2019hko,Choi:2019miv}. However, these decoupled subsectors are not as well-studied as their counterpart in $\mathcal{N}=4$ SYM, which is known as the Spin Matrix theory \cite{Harmark:2007px,Harmark:2014mpa,Harmark:2019zkn,Baiguera:2020jgy,Baiguera:2020mgk,Baiguera:2021hky,Baiguera:2022pll}. A major challenge is that the SCFTs in $D=5,6$, and especially the $6d$ $\mathcal{N}=(2,0)$ theory, are far from being completely understood as they lack weak coupling descriptions \cite{Chang:2018xmx}, although relevant works to construct Lagrangian descriptions can be found in \cite{Lambert:2019diy}.
Despite these challenges, our analysis indicates that the decoupled subsectors of SCFTs in $D=5,6$ provide a natural framework for describing the holographic duals to asymptotically flat EMMD black holes.
Ultimately, these models propose a novel algorithm: utilizing higher-dimensional AdS/CFT embeddings to decode the microscopic states of flat or non-AdS black holes, thereby providing a concrete example of holography beyond the standard AdS/CFT correspondence.

Analytic solutions for black objects in higher dimensions are more readily available in asymptotically flat spacetimes. Notable examples include the Myers–Perry black holes with the maximal number of independent rotations \cite{Myers:1986un}, the Cvetic–Youm solution incorporating U$(1)$ charges in arbitrary dimensions \cite{Cvetic:1996dt}, as well as solutions with nontrivial topologies such as black rings \cite{Emparan:2001wn,Emparan:2006mm,Bena:2014rea} and black saturns \cite{Elvang:2007rd}. For comprehensive reviews, we refer to \cite{Obers:2008pj}. These nontrivial configurations may give rise to even richer geometric structures in the near-EVH regime \cite{Ghodsi:2014fta,Golchin:2013con}.

Unlike AdS$_5$ black hole, whose near horizon limits are pinched AdS$_3$/BTZ geometry \cite{Sheikh-Jabbari:2011sar,deBoer:2010ac, deBoer:2011zt}, the near EVH limits of AdS$_{D=6,7}$ black holes are of the following forms 
\begin{equation}\label{eq:decoupled-ansatz}
	ds_{D}^2 = \epsilon^2 \Omega \times ds_{\text{EMMD}_{D-2}}^2 + ds_{S^2}^2 \,,
\end{equation}
where $\Omega$ is a conformal factor. 
These kinds of geometries are similar to the recently found M$p$T geometries which are generalizations of the non-relativistic Newton-Cartan geometry to make the D$p$-brane actions have well-defined non-relativistic limits \cite{Blair:2023noj,Blair:2024aqz}.
The form of decoupled geometry \eqref{eq:decoupled-ansatz} in the near-EVH limits of AdS$_{6,7}$ enlightens that we can study them from the perspectives of non-relativistic strings and branes. 
They might provide a non-perturbative configurations in non-relativistic string theory with finite temperature. 

The EMMD geometries emerging in the near-EVH limits of AdS$_{D}$ black holes are specific to particular gauged supergravity models, whereas AdS$_2$ throats universally appear in the near-horizon region of near-extremal black holes. Although finite-energy excitations break the strict AdS$_2$ boundary conditions, their dynamics are universally captured by Jackiw–Teitelboim (JT) gravity—a nearly AdS$_2$ theory proposed to be dual to the Sachdev–Ye–Kitaev (SYK) model \cite{Almheiri:2014cka,Engelsoy:2016xyb,Cvetic:2016eiv,Maldacena:2016upp,Maldacena:2016hyu}. This framework accounts for the linear term $S_1 T$ in the low-temperature entropy expansion $S = S_0 + S_1 T$; see \cite{Sarosi:2017ykf,Mertens:2022irh} for pedagogical treatments. Extensions to three dimensions have also been explored, providing a higher-dimensional analogue of the near-AdS$_2$/SYK correspondence \cite{Cotler:2018zff}. These developments have further been applied to study the dynamics of near-BPS AdS$_5$ black holes and to identify mass gaps \cite{Boruch:2022tno}.
In our context, the EMMD black holes serve as novel low-energy effective geometries in the infrared, characterized by higher power-law scalings $S \sim T^{D-4}$. They may similarly be understood as gravitational duals to strongly coupled low-energy sectors, much like higher-dimensional generalizations of the SYK model capturing nontrivial infrared dynamics.

\section*{Acknowledgement}

We thank Yan Liu, Hong Lv, Sanjaye Ramgoolam, Ziqi Yan for useful discussions and also anonymous referee for useful comments on the draft. 
Y.L. also thanks Nordita, Eurostrings 2025 and Niels Bohr Institute where this work is in progress. 
Y.L. is supported by a Project Funded by the Priority Academic Program Development of Jiangsu Higher Education Institutions (PAPD) and by National Natural Science Foundation of China (NSFC) No.12305081 and the international collaboration grant between NSFC and Royal Society No.W2421035.

\appendix
\section{Review of EMMD gravity}
\label{appendix:EMMD}

In this section, we provide a brief review of the Einstein–Maxwell–Maxwell–Dilaton gravity model investigated in \cite{Lu:2013eoa,Lu:2025eub}.
The Lagrangian of the $D$-dimensional supergravity under investigation is given by
\begin{equation}\label{eq:EMMD-action}
	\frac{1}{e} \mathcal{L} = R - \frac{1}{2} (\partial \phi)^2 - \frac{1}{4} e^{a_1 \phi} F_1^2 - \frac{1}{4} e^{a_2 \phi} F_2^2 - V(\phi),
\end{equation}
where $a_1$ and $a_2$ are dilaton coupling constants.
To obtain analytic solutions in this supergravity theory, it is convenient to parameterize the couplings as
\begin{equation}
	a_1^2 = \frac{4}{N_1} - \frac{2(D-3)}{D-2}, \qquad
	a_2^2 = \frac{4}{N_2} - \frac{2(D-3)}{D-2},
\end{equation}
with $N_i$ being positive integers in the context of supergravity.
Our focus lies particularly on black hole solutions that satisfy
\begin{equation}
	a_1 a_2 = - \frac{2(D-3)}{D-2}.
\end{equation}
Under this condition, analytic solutions carrying two independent charges can be constructed, as shown in \cite{Lu:2013eoa}.

General solutions to the field equations derived from the action \eqref{eq:EMMD-action} take the form
\begin{align}\label{eq:metric-EMMD}
	\begin{split}
ds^2 &= -(H_1^{N_1} H_2^{N_2})^{- \frac{D-3}{D-2}} f dt^2 + (H_1^{N_1} H_2^{N_2})^{\frac{1}{D-3}} \left(\frac{dr^2}{f} +r^2 d\Omega_{D-2}^2 \right) \\
A_1 &= \frac{\sqrt{N_1} c_1}{s_1} \frac{dt}{H_1}, \qquad A_2 = \frac{\sqrt{N_2} c_2}{s_2} \frac{dt}{H_2} \\
\phi& =\frac{1}{2} N_1 a_1 \ln H_1 + \frac{1}{2} N_2 a_2 \ln H_2, \qquad f = 1- \frac{\mu}{r^{D-3}} \\
H_1 & = 1+ \frac{\mu s_1^2}{r^{D-3}} , \qquad H_2 = 1+ \frac{\mu s_2^2}{r^{D-3}}  \,,
	\end{split}
\end{align}
subject to the constraints
\begin{equation}
	N_1 a_1 +N_2 a_2 =0, \qquad N_1 +N_2 = \frac{2(D-2)}{D-3} \,.
\end{equation}
In four dimensions ($D=4$), the possible solutions include $(N_1, N_2) = (2,2),\ (1,3),\ (3,1)$, while in three dimensions ($D=5$) the only admissible pairs are $(N_1, N_2) = (2,1),\ (1,2)$.
We now examine each of these cases in turn.

\subsection*{$D=4$ geometry with $(N_1,N_2)=(2,2)$}

This symmetric choice of parameters, with $a_1 = -a_2 = 1$, yields a particularly simple and elegant solution. 
Defining the charge parameters as
\begin{equation}
	q_1= \mu s_1^2, \qquad q_2 =\mu s_2^2\,,
\end{equation}
the metric in \eqref{eq:metric-EMMD} simplifies to 
\begin{equation}
	ds^2 = - \frac{r(r-\mu)}{(r+q_1)(r+q_2)} dt^2 + \frac{(r+q_1)(r+q_2)}{r(r-\mu)} dr^2 + (r+q_1)(r+q_2) d\Omega_2^2 \,.
\end{equation}
To reveal its underlying conformal structure, we can factor the metric. 
Assuming without loss of generality that $q_2 - q_1 = \mathfrak{q} > 0$, it can be recast into the form 
\begin{equation}\label{eq:EMMD-4D-N1=N2}
ds^2 = \frac{r+q_2}{r+q_1} \left[ 
- \frac{(r-\mu)r}{ (r+q_2)^2} dt^2 + \frac{(r+q_1)^2}{r(r-\mu)} dr^2 + (r+q_1)^2 d\Omega_2^2
\right] \,.
\end{equation}
This conformal representation highlights how the solution interpolates between different geometries but with the same asymptotics, with the prefactor playing a crucial role.
 
\subsection*{$D=4$ geometry with $(N_1,N_2)=(1,3)$}

Due to the symmetry in the indices of $N_1$ and $N_2$, it is sufficient to analyze the $(1,3)$ case, as the $(3,1)$ solution is its direct counterpart. 
The corresponding dilaton couplings are fixed as
\begin{equation}
	a_1 = \sqrt{3}, \qquad a_2 = -\frac{1}{\sqrt{3}} \,.
\end{equation}
In this asymmetric configuration, the metric \eqref{eq:metric-EMMD} becomes
\begin{equation}
	ds^2 = - \frac{r(r-\mu)}{\sqrt{(r+q_1)(r+q_2)^3}} dt^2 + \frac{\sqrt{(r+q_1)(r+q_2)^3}}{r(r-\mu)} dr^2 +\sqrt{ (r+q_1)(r+q_2)^3} d\Omega_2^2 \,.
\end{equation}
Expressing this metric in a conformally rescaled frame provides a clearer geometric interpretation:
\begin{equation}
	ds^2 = \sqrt{\frac{r+q_1}{r+q_2}} \left[
-\frac{r(r-\mu)}{(r+q_1)(r+q_2)} dt^2 + \frac{(r+q_2)^2 dr^2}{r(r-\mu)} + (r+q_2)^2	 d\Omega_2^2
	\right] \,.
\end{equation}

\subsection*{$D=5$ geometry with $(N_1,N_2)=(1,2)$}

We now turn to a five-dimensional example characterized by $(N_1, N_2) = (1, 2)$. The associated dilaton couplings are given by
\begin{equation}
	a_1 = \sqrt{\frac{8}{3}} , \qquad a_2 = -\sqrt{\frac{2}{3}} \,.
\end{equation}
The full metric in this case reads
\begin{align}
	\begin{split}
ds^2 &= - \frac{r^2(r^2-\mu) dt^2}{(r^2+q_1)^{\frac{2}{3}}  (r^2+q_2)^{\frac{4}{3}} } + \frac{
(r^2+q_1)^{\frac{1}{3}}  (r^2+q_2)^{\frac{2}{3}}
}{r^2- \mu} dr^2 + (r^2+q_1)^{\frac{1}{3}} (r^2+q_2)^{\frac{2}{3}} d\Omega_3^2\,.
	\end{split}
\end{align}
A clearer picture of the geometry emerges after performing the coordinate transformation
\begin{equation}
	x^2 =r^2+q_1, \qquad \mathfrak{q} =q_2-q_1 >0 \,.
\end{equation}
In these new coordinates, the metric simplifies to the following conformally flat form:
\begin{align}\label{eq:EMMD-5d}
	\begin{split}
ds^2 &= \frac{(x^2+\mathfrak{q})^{\frac{2}{3}}}{x^{\frac{4}{3}}} \left[
-\frac{(x^2-q_1)(x^2-q_1-\mu) dt^2}{(x^2+\mathfrak{q})^2} + \frac{x^4 dx^2}{(x^2-q_1)(x^2-q_1-\mu) } + x^2 d\Omega_3^2
\right]\,.
	\end{split}
\end{align}
This final form reduces to the standard five dimensional RN-black hole by taking $\mathfrak{q}=0$.

\bibliographystyle{JHEP}
\bibliography{qec}
 
\end{document}